\documentclass[12pt,letterpaper]{article}
\usepackage[left=1in,right=1in,top=1in,bottom=1in]{geometry}
\usepackage{amsmath,amssymb,amsfonts,bm}
\usepackage{graphicx,wrapfig}
\usepackage{multirow}
\usepackage{verbatim}
\usepackage{appendix}
\usepackage{rotating}

\graphicspath{{./images/}}

\newcommand{\bea}{\begin{eqnarray}}
\newcommand{\eea}{\end{eqnarray}}
\newcommand{\be}{\begin{equation}}
\newcommand{\ee}{\end{equation}}
\newcommand{\bse}{\begin{subequations}}
\newcommand{\ese}{\end{subequations}}
\newcommand{\mb}{\mathbf}

\newcommand{\wt}{\widetilde}
\newcommand{\wh}{\widehat}

\newcommand{\ol}{\overline}

\newcommand{\eg}{\emph{e.g.}}
\newcommand{\ie}{\emph{i.e.}}
\newcommand{\cf}{\emph{cf.}}

\newcommand{\Z}{{\mathbb Z}}
\newcommand{\R}{{\mathbb R}}
\newcommand{\C}{{\mathbb C}}

\newcommand{\Li}{{\rm Li}}
\newcommand{\cp}{{\mathbb{CP}}}

\newcommand{\Tr}{{\rm Tr \,}}
\renewcommand{\Re}{{\rm Re}}
\renewcommand{\Im}{{\rm Im}}
\newcommand{\bs}{\backslash}
\newcommand{\pd}{\partial}

\newcommand{\CA}{\mathcal{A}}

\newcommand{\CC}{\mathcal{C}}

\newcommand{\CH}{\mathcal{H}}

\newcommand{\CK}{\mathcal{K}}
\newcommand{\CL}{\mathcal{L}}
\newcommand{\CM}{\mathcal{M}}
\newcommand{\CN}{\mathcal{N}}
\newcommand{\CO}{\mathcal{O}}
\newcommand{\CP}{\mathcal{P}}

\newcommand{\CR}{\mathcal{R}}
\newcommand{\CS}{\mathcal{S}}

\newcommand{\CZ}{\mathcal{Z}}

\begin{document}

\begin{center} {\Large \bf 3d Superconformal Theories from Three-Manifolds} \end{center}
\vspace{0in}

\begin{center} Tudor Dimofte 
\end{center}

\noindent \hspace{1in}\emph{Institute for Advanced Study, Einstein Dr., Princeton, NJ 08540}

\noindent \hspace{1in}\emph{Trinity College, Cambridge CB2 1TQ, UK}

\tableofcontents
\noindent\hrulefill

\vspace{.7in}

We review here some aspects of the 3d $\CN=2$ SCFT's that arise from the compactification of M5 branes on 3-manifolds.
The program to systematically describe these theories and their properties began in a series of papers \cite{DGG, CCV, DGG-index}, inspired by earlier physical studies \cite{DGH, Yamazaki-3d, DG-Sdual}, and has since been extended and clarified in \cite{DGG-Kdec, DGV-hybrid, Cordova-tangles, FGSS-AD, CJ-S3, LY-S2}, among other works.

Part of the ``3d-3d correspondence'' includes an analogue of the AGT relation \cite{AGT, AGGTV, DGOT, HH-S4b} and the index-TQFT relation of \cite{Rastelli-2dQFT, Rastelli-qYM, GRR-bootstrap}, discussed in much of the rest of this volume. Recall that for theories of class $\CS$, \ie\ 4d $\CN=2$ theories $T_K[C]$ obtained by wrapping $K$ M5 branes on $C$, one expects
\be \label{3d:AGT-intro}
\begin{array}{ccc}
\text{Partition function of $T_K[C]$ on \underline{\quad\;}} &&   \text{Partition function of \underline{\quad\;} on $C$} \\\hline
 S^4_b   &=&  \text{Liouville theory} \\
 \R^4_\epsilon \simeq D^4_b &=& \text{Liouville theory (conformal block)} \\
 S^3\times_q S^1 &=& \text{$q$-Yang-Mills or generalizations}
\end{array}
\ee
The basic logic is that one takes the 6d geometry supporting the (2,0) theory on the worldvolume of the M5 branes to be of the form $X\times C$, where $X$ is one of the geometries in the left column of \eqref{3d:AGT-intro}; then compactifying first on $C$ leads to $T_K[C]$ on $X$, whereas compactifying first on $X$ should lead to some other theory (the right column of \eqref{3d:AGT-intro}) on $C$. Similarly, if we denote by $T_K[M]$ the effective 3d field theory obtained from wrapping $K$ M5 branes on $M$, we expect%
\footnote{Here $S^3_b$ denotes a ``squashed'' 3-sphere with ellipsoidal metric. It is also useful to note that complex $SL(K,\C)$ Chern-Simons theory has \emph{two} coupling constants or levels $(k,\sigma)$, one quantized and the other continuous, \cf\ Sections \ref{3d:sec:quant}--\ref{3d:sec:AGT}. It is only the quantized level that is being fixed in \eqref{3d:3d3d-intro}. The general pattern following from work of \cite{CJ-S3} is that $T_K[C]$ on a squashed Lens space $L(k,1)_b$ is equivalent to $SL(K,\C)$ Chern-Simons at level~$k$.}
\be \label{3d:3d3d-intro}
\begin{array}{ccc}
\text{Part'n function of $T_K[C]$ on \underline{\quad\;}} &&
 \text{Part'n function of \underline{\quad\;} on $M$} \\\hline
 S^3_b  &=&  \text{$SL(K,\C)$ Chern-Simons at level $k=1$ \cite{DGG, CJ-S3}} \\
 S^2\times_q S^1 &=& \text{$SL(K,\C)$ Chern-Simons at level $k=0$ \cite{DGG-index, LY-S2}} \\
 \R^2\times_q S^1 &=& \text{holomorphic sector of $SL(K,\C)$ CS \cite{Wfiveknots, DGH, BDP-blocks}} \\\hline
 \text{SUSY vacua on $\R^2\times S^1$} &=& \text{flat $SL(K,\C)$ connections on $M$ \cite{Wfiveknots, DGH}\,.}
\end{array}
\ee

The 3d-3d and 2d-4d correspondences fit very nicely together when $M$ has a boundary. We will describe in Section \ref{3d:sec:6d} that when $\pd M$ is nontrivial, the theory $T_K[M]$ is best interpreted as a boundary condition or domain wall for the 4d $\CN=2$ theory $T_K[\pd M]$ \cite{DGG-defects, DGG, DGV-hybrid}.
This has some natural implications for partition functions. For example, if $M$ has two distinct boundaries of the same type, $\pd M=C\sqcup C$, then $T_K[M]$ describes a domain wall in the 4d theory $T_K[C]$. In turn, the 3d partition functions of $T_K[M]$ on a space $Y$ (from the right column of \eqref{3d:3d3d-intro}) should \emph{act} on the 4d partition functions of $T_K[C]$ on a half-space $X$ with $\pd X = Y$. Examples of this type and others have been explored in \cite{HLP-wall, Yamazaki-3d, DG-Sdual, TV-6j, Gang-index}, and we will elaborate a bit further on them in Section \ref{3d:sec:AGT}. Similar ideas about domain walls also constituted a major ingredient in the recent 4d-2d correspondence of \cite{GGP-4d}.



The current successes of the ``3d-3d'' program include a systematic prescription for associating theories $\wt T_K[M]$ to a wide class of 3-manifolds $M$ with boundary \cite{DGG, CCV, DGG-Kdec}, which we discuss in Sections \ref{3d:sec:topdown}--\ref{3d:sec:bottomup}. Sometimes the theories $\wt T_K[M]$ only contain a subsector%
\footnote{\label{3d:foot:sub}To be precise: after compactification on $S^1$, the subsectors only contain SUSY vacua corresponding to irreducible $SL(K,\C)$ flat connections on $M$, with given boundary conditions, rather than \emph{all} flat connections as prescribed by \eqref{3d:3d3d-intro}. The relation between these subsectors and the ``full'' $T_K[M]$ began to be analyzed in~\cite{CDGS}.} %
of the full theory $T_K[M]$ of $K$ M5 branes on $M$; though in special cases one does recover the full $T_K[M]$.
In particular, one recovers the full $T_K[M]$ when $M$ is a 3-manifold encoding a duality domain wall in a 4d $\CN=2$ $T_K[C]$, as long as $\chi(C)<0$. We will revisit this subtlety in Section \ref{3d:sec:framed}; in the following we drop the tilde on $\wt T_K[M]$ to simplify notation.

The main technique of \cite{DGG, CCV, DGG-Kdec} is to triangulate the manifold, cutting it up into tetrahedra, and then to ``glue'' $T_K[M]$ together from elementary 3d theories $T_\Delta$ associated to the tetrahedron pieces. One obtains this way an abelian Chern-Simons-matter theory --- a theory of ``class $\CR$'' --- that flows to the desired SCFT $T_K[M]$ in the infrared.
Quite beautifully, different triangulations of $M$ lead to different UV Chern-Simons-matter theories that flow to the \emph{same} $T_K[M]$. In other words, the UV theories are related by a generalized 3d mirror symmetry. The 3d-3d program therefore leads to the \emph{geometric} classification of a huge subset of abelian 3d mirror symmetries.

Mathematically, the study of 3-manifold theories based on triangulations has led to the new concepts of ``framed'' 3-manifolds and moduli spaces of ``framed'' flat connections on them \cite{DGG-Kdec, DGV-hybrid}. They generalize the framework of \cite{FG-Teich} for studying higher Teichm\"uller theory on 2d surfaces --- which in turn played a central role in the 2d-4d explorations of Gaiotto, Moore, and Neitzke, \cf\ \cite{GMNII, GMN-snakes}.

Despite many exciting achievements, there is still much to develop in the 3d-3d program. One interesting direction of study would be to find \emph{nonabelian} UV descriptions for theories $T_K[M]$, dual to the abelian ones that come from triangulations.%
\footnote{In a few examples, nonabelian duals are already known: the basic tetrahedron theory has an $SU(2)$ dual discussed in \cite{JY-appetizer}; and the theory for the basic S-duality wall in 4d $\CN=2$ $SU(2)$ theory with $N_f=4$ (associated to the manifold in Figure \ref{3d:fig:wallgeoms}b below) has an $SU(2)$ dual found in \cite{TV-6j}. Some basic ideas about smooth gluing were also discussed in \cite{Yamazaki-3d}.} %
This may come from cutting manifolds into simpler pieces along smooth surfaces (rather than sharp tetrahedron boundaries, which have edges and corners), much as was done for cutting 2d surfaces in \cite{Gaiotto-dualities}.
Such smooth cutting and gluing should provide the construction of $T_K[M]$ for general closed 3-manifolds as well, and may circumvent the difficulties with irreducible flat connections and subsectors (\cf\ Footnote \ref{3d:foot:sub}) encountered so far. Finally, while computations of sphere partition functions and indices of $T_K[M]$ are easy and accessible, it would be extremely interesting to analyze the actual Q-cohomology of the space of BPS states of a theory $T_K[M]$ on (say) $S^2\times \R$. This would have immediate applications to the categorification of quantum 3-manifold invariants, along the lines of~\cite{GIKV, Wfiveknots}.


\section{The 6d setup}
\label{3d:sec:6d}

Before discussing methods to construct $T_K[M]$, let's first try to understand exactly what it \emph{means} to associate a 3d $\CN=2$ theory to an oriented 3-manifold $M$, and what properties the theory should have.

One way to think about this is to start in 11-dimensional M-theory, wrapping $K$ M5 branes on $M\times \R^3$. If we want to preserve supersymmetry we must make sure that $M$ is a supersymmetric cycle. Taking the ambient 11-dimensional geometry to be a cotangent bundle $T^*M\times \R^5$ (with $M$ its zero-section), we can preserve at least four supercharges.%
\footnote{The counting goes as follows. First, the cotangent bundle $T^*M$ is a noncompact Calabi-Yau manifold. M-theory on a generic Calabi-Yau background  preserves eight supercharges (\cf\ \cite{CCDF-CY, PT-CY}). An M5 brane wrapping a special Lagrangian cycle in the Calabi-Yau (such as the zero-section $M$ in $T^*M$) is half-BPS, and preserves four of the eight supercharges.} %
If we subsequently decouple gravity, taking a field-theory limit on the M5 branes, and flow to low energy so that fluctuations along $M$ can be neglected, we expect to obtain a 3-dimensional $\CN=2$ theory on $\R^3$. In the far infrared, the theory generically hits a superconformal fixed point, which we might call $T_K[M]$.

In this brane construction, the starting metric on $M$ might be chosen arbitrarily. All the details of the metric enter (as couplings) into an effective field theory on $\R^3$.
 However, in the process of flowing to the infrared the metric is expected to ``uniformize,'' acquiring constant curvature.%
\footnote{See, \eg, the supergravity solutions of \cite{GKW-5branes} involving special Lagrangian 3-cycles. For the analogous compactifications on 2d surfaces, the flow of the metric to constant curvature was analyzed in~\cite{ABBR-uniform}.} %
Correspondingly, renormalization flow washes away most of the coupling dependence in the effective theory on $\R^3$.
Most topological 3-manifolds admit a metric with constant negative curvature \cite{thurston-1982}, \ie\ a hyperbolic metric, and they are the ones we'll be interested in.%
\footnote{Notable exceptions include spheres, tori, lens spaces, and more general Seifert-fibered manifolds, which have the structure of an $S^1$ fibration over a surface. The 3d theories resulting from compactification on such manifolds are qualitatively different from the hyperbolic case. For example, compactification on a 3-torus yields $\CN=8$ SYM in 3d, while compactification on the 3-sphere yields a gapped theory that breaks SUSY.} %
Moreover, if $M$ is closed, the hyperbolic metric is unique \cite{mostow-1973}. In this case, $T_K[M]$ is indeed expected to be a superconformal theory, which depends only on the \emph{topology} of $M$, has no flavor symmetry, and admits no (obvious) marginal deformations. Just as the hyperbolic structure on $M$ is rigid, we might say that $T_K[M]$ is rigid.

We may also understand $T_K[M]$ directly in field theory. The 6d theory on $K$ M5 branes is the $(2,0)$ SCFT with Lie algebra $A_{K-1}$.
It must be topologically twisted along $M$ in order to preserve supersymmetry. (In general, the required topological twist is prescribed by the normal geometry of the supersymmetric cycle $M \subset T^*M$ \cite{BSV-Dtop}; but in this case the choice is unique.)
In particular, the $SO(3)_E$ part of the Lorentz group corresponding to $M$ is twisted by an $SO(3)_R$ subgroup of the $SO(5)_R$ R-symmetry group (\cf\ \cite{Wfiveknots, DGH}). The unbroken R-symmetry is the commutant of $SO(3)_R\subset SO(5)_R$, namely $SO(2)_R\simeq U(1)_R$, as appropriate for an $\CN=2$ theory in 3d. We again are welcome to choose any metric on $M$ that we want. In the UV, the effective field theory on $\R^3$ will depend on the metric, but after flowing to the IR one hopes to obtain an SCFT that does not.

This is all entirely analogous to compactification of $K$ M5 branes, or the $A_{K-1}$ (2,0) theory, on 2d surfaces $C$. In that case, the IR theory $T_K[C]$ (a theory of ``class $\CS$'') depends on the conformal class of a metric on $C$, which is equivalent to a choice of hyperbolic metric. In contrast to 3d, the hyperbolic metric on a closed surface allows continuous deformations, and the 4d $\CN=2$ theory $T_K[C]$ has corresponding exactly marginal gauge couplings \cite{Ga}.

The story becomes much more interesting, and in many ways much more manageable, if we allow $M$ to have defects and boundaries.

Codimension-two defects placed along knots in $M$ add flavor symmetry to $T_K[M]$. In the 6d $A_{K-1}$ (2,0) theory, there are different types of ``regular'' defects, labelled by partitions of $K$, and carrying various subgroups of $SU(K)$ as their flavor symmetry \cite{Gaiotto-dualities}%
\footnote{Also described in Sec. 3.1--3.2 of \emph{Families of $\CN=2$ field theories} by D. Gaiotto.}. %
In M-theory, each regular defect along a knot $\CK\subset M$ comes from a stack of $K$ or fewer ``probe'' M5 branes that wrap the noncompact supersymmetric 3-cycle $N^*\CK \subset T^*M$ (the conormal bundle of $\CK$) as well as $\R^3$. The flavor symmetry can be understood as arising from the symmetry group of the probes. In the presence of a defect, the hyperbolic metric on $M$ acquires a cusp-like singularity, \cf\ \cite{GaiottoMaldacena}.

\begin{figure}[htb]
\centering
\includegraphics[width=5.5in]{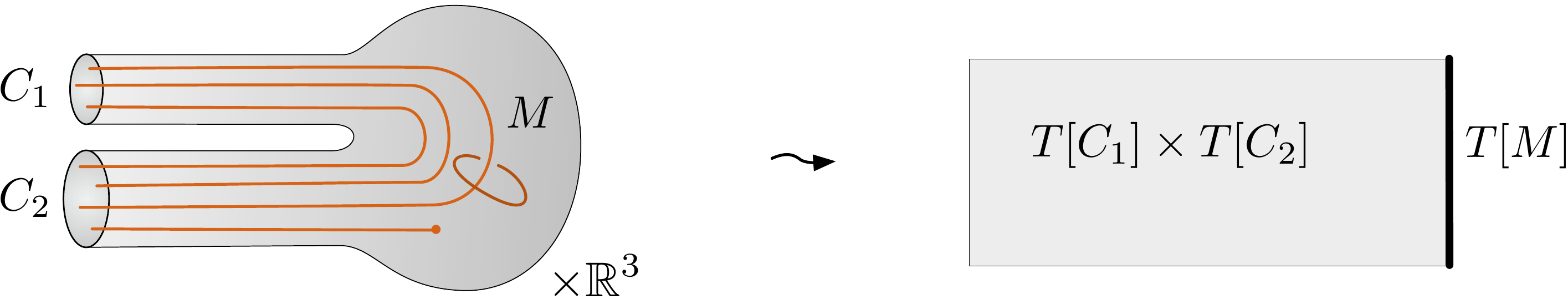}
\caption{\small Compactifying M5's on a 3-manifold with asymptotic boundaries to obtain a boundary condition $T[M]$.} 
\label{3d:fig:regSW}
\end{figure}

\begin{figure}[htb]
\centering
\includegraphics[width=5.5in]{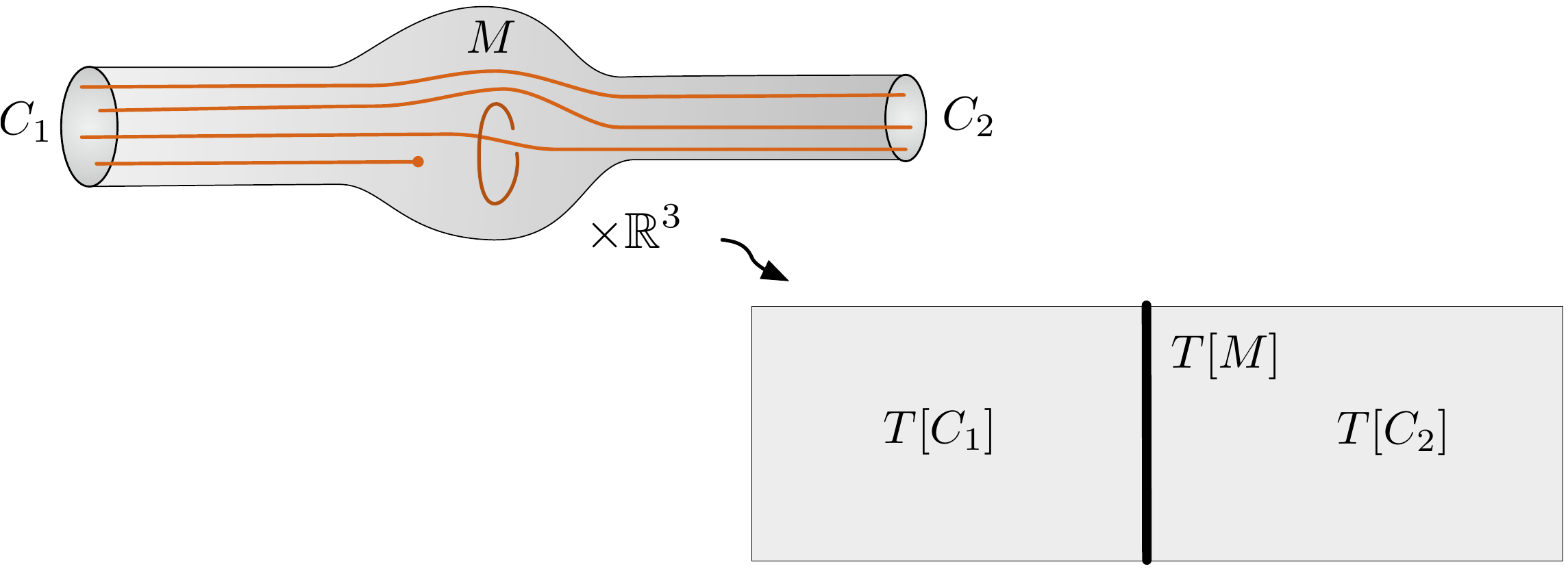}
\caption{\small Re-interpreting the boundary condition of Figure \ref{3d:fig:regSW} as a domain wall.} 
\label{3d:fig:domSW}
\end{figure}

In order to add boundaries to $M$, we must be somewhat more creative, since M5 branes cannot end. Alternatively, the $(2,0)$ theory does not admit ordinary supersymmetric boundary conditions because it is chiral. We create boundaries for $M$ at ``infinity'' by allowing asymptotic regions that look like $\R_+\times C$ for some surface $C$ (Figure \ref{3d:fig:regSW}). Then $M$ is no longer compact. Wrapping M5 branes on $M$ leads not to an isolated 3d theory but to a half-BPS superconformal boundary condition (preserving 3d $\CN=2$ SUSY) for the 4d theory $T_K[C]$. We might call this boundary condition $T_K[M]$. If $M$ has multiple asymptotic regions with cross-sections $C_i$, then $T_K[M]$ is a common boundary condition for a product of theories $T_K[C_i]$, which do not interact with each other in the 4d bulk; equivalently, $T_K[M]$ can be thought of as a half-BPS domain wall between one subset of 4d theories $\prod_{i< I} T_K[C_i]$ and its complement $\prod_{i\geq I} T_K[C_i]$ (Figure \ref{3d:fig:domSW}).
Note that defects in $M$ (orange lines in the figures) may enter asymptotic regions, where they look like punctures in the surfaces $C_i$.

In the presence of asymptotic boundaries $C_i$, the hyperbolic metric on $M$ is no longer rigid. It depends (at least) on a choice of hyperbolic structure for each surface $C_i$, \ie\ on a choice of boundary conditions. This choice, of course, parametrizes the bulk couplings of $\prod_i T_K[C_i]$.

We can try to transform the boundary condition $T_K[M]$ into a stand-alone 3d $\CN=2$ theory by decoupling the 4d bulk theories $\prod_i T_K[C_i]$. However, there is no unique way to do this. Suppose, for example, that there's just a single boundary $C$. Working with a nonabelian SCFT $T_K[C]$, we attain a (non-canonical) weak-coupling limit by adjusting the hyperbolic metric on $C$ so as to stretch it into pairs of pants connected by long, thin tubes \cite{Gaiotto-dualities}. There is a weakly coupled $SU(K)$ gauge group in $T_K[C]$ associated to each tube. In the limit of infinite stretching, we may hope to leave behind a 3d theory $T_K[M,\mb p]$, labelled by the chosen pants decomposition $\mb p$ of $C$. $T_K[M,\mb p]$ should have a residual $SU(K)$ flavor symmetry for every stretched tube (which would get gauged in re-coupling to a 4d bulk).

\begin{figure}[htb]
\centering
\includegraphics[width=4.2in]{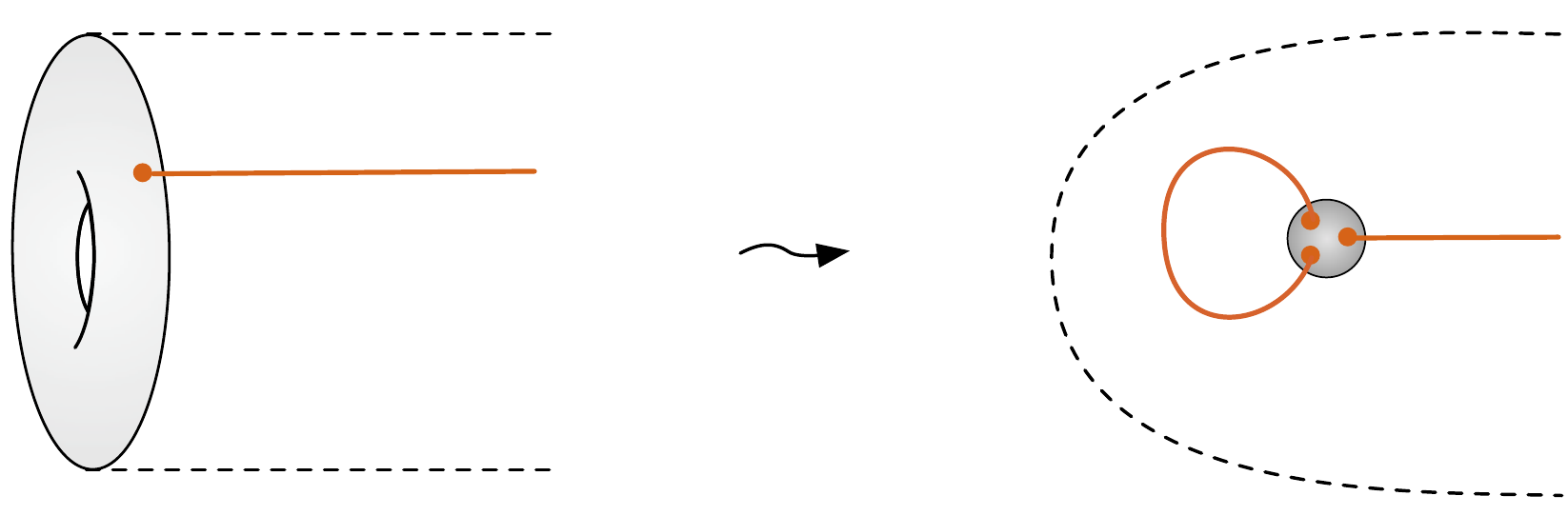}
\caption{\small Shrinking a pants decomposition of a one-punctured torus into a network of defects.}
\label{3d:fig:shrink}
\end{figure}

We may represent this 4d--3d decoupling geometrically by ``shrinking'' $C$ to a trivalent network of maximal codimension-two defects, as dictated by the pants decomposition $\mb p$ (Figure \ref{3d:fig:shrink}).%
\footnote{This ``shrinking'' procedure turns parts of $M$ that look like $S^1\times \R\times \R_+$ (\ie\ the neighborhoods of tubes) into defects. An identical setup was used to create defects in \cite{Gaiotto-dualities}.} %
This effectively compactifies $M$. The trivalent junctures of defects survive as asymptotic regions of $M$ with the cross-section of a 3-punctured sphere. Thus, the theory $T_K[M,\mb p]$ is still potentially coupled to a collection of 4d ``trinion'' theories, and this coupling takes a little extra work to undo. For example, in the case $K=2$, the trinion theory just consists of four free hypermultiplets, coupled to the 3d boundary theory by superpotentials.
One can adjust bulk parameters to make some of these hypermultiplets very massive. This is discussed in \cite{DGG} and especially \cite{DGV-hybrid}.

Alternatively, we can move onto the 4d Coulomb branch of $T_K[C]$ and flow to the IR. Then $T_K[C]$ is a Seiberg-Witten theory, with some abelian gauge symmetry $U(1)^d$. The electric-magnetic duality group is $Sp(2d,\Z)$.
We decouple the Seiberg-Witten theory by choosing an electric-magnetic duality frame $\Pi$, and adjusting parameters and moduli so that all the electric gauge couplings in that frame become weak. Again, a little more is needed to decouple BPS hypermultiplets. In the end, we obtain a purely 3d theory $T_K[M,\Pi]$ with $U(1)^d$ flavor symmetry left over from the bulk gauge group. We will usually represent the manifold giving rise to $T_K[M,\Pi]$ as simply having its asymptotic region $C\times \R_+$ cut off at finite distance.


\subsection{Duality walls}
\label{3d:sec:duality}

A very simple application of the above constructions is to represent duality walls for 4d $\CN=2$ theories of class $\CS$ by 3d geometries \cite{Yamazaki-3d, DG-Sdual, DGV-hybrid}. To this end, we take $M=\R\times C$ for some punctured surface $C$. In other words, $M$ has two asymptotic boundaries $C$. The punctures of $C$ just become defects running the entire ``length'' of $M$. Naively, $T_K[M]$ just becomes a trivial domain wall between two copies of $T_K[C]$. However, we can make it look non-trivial by taking different decoupling limits on the two ends.

\begin{figure}[htb]
\centering
\includegraphics[width=5.7in]{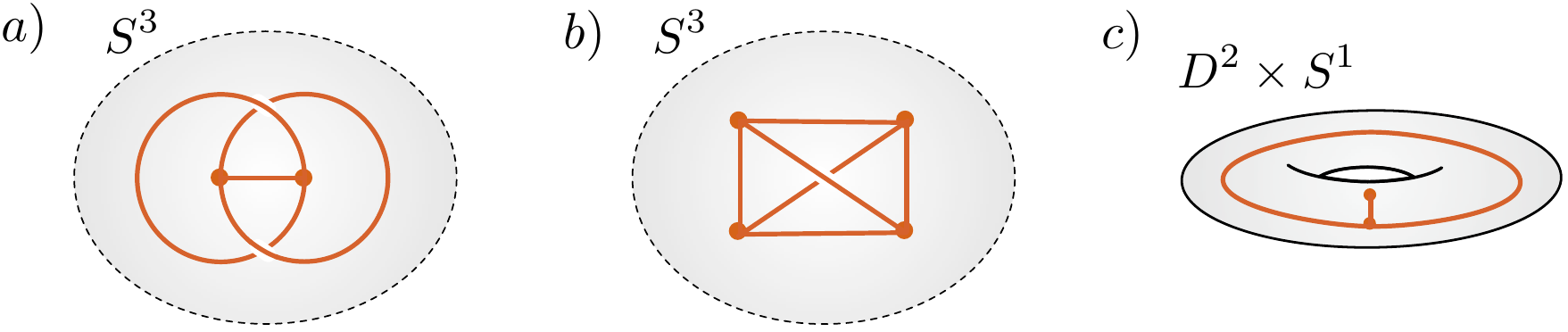}
\caption{\small Geometries representing various S-duality and RG walls: a) a Hopf network of defects in $S^3$; b) a tetrahedral network of defects in $S^3$; and c) a network of defects in a solid torus, corresponding to a particular pants decomposition and connecting to a puncture on the boundary.}
\label{3d:fig:wallgeoms}
\end{figure}

For example, if we work with $T_K[C]$ as a UV SCFT, we can take two different weak-coupling limits corresponding to pants decompositions $\mb p,\mb p'$.
The 3d theory $T_K[M,\mb p,\mb p']$ that is left behind is the theory of an S-duality domain wall. For example, if $C$ is a punctured torus (with a minimal puncture), then $T_K[C]$ is 4d $\CN=2^*$ theory with gauge group $SU(K)$. Letting $\mb p$ and $\mb p'$ shrink the A and B-cycles of the punctured torus, respectively, we should obtain the S-duality wall whose 3d theory is usually called (mass-deformed) $T[SU(K)]$ \cite{GW-Sduality}.

As discussed above, we can represent decoupling limits geometrically by shrinking appropriate legs/tubes of $C$ to defects at the two ends of $M$, so that we obtain a compact manifold $M_{\mb p,\mb p'}$ with a trivalent network of defects. In the case of S-duality for a one-punctured torus, the resulting manifold is a ``Hopf network'' of defects in $S^3$, shown in Figure \ref{3d:fig:wallgeoms}a. 
By using the methods of Section \ref{3d:sec:bottomup}, its 3d theory was shown in \cite{DGV-hybrid} to be equivalent to $T[SU(2)]$ (for $K=2$). Similarly, if we take $C$ to be a four-punctured sphere (with appropriate minimal/maximal punctures) and set $\mb p,\mb p'$ to correspond to its ``$s$ and $t$ channel'' decompositions, we get the basic S-duality for $\CN=2$ SQCD with $N_f=2K=2N_c$. The 3d geometry for the duality wall is shown in Figure \ref{3d:fig:wallgeoms}b; its associated 3d theory appeared in \cite{TV-6j, DGV-hybrid}.

We can also put theories $T_K[C]$ on their Coulomb branch, and choose decoupling limits $\Pi,\Pi'$ at the two ends of $M$ that are appropriate for Seiberg-Witten theory. The theory $T_K[M,\Pi,\Pi']$ becomes a ``Seiberg-Witten duality wall'' that implements abelian IR dualities. The simplest such walls (involving duality for gauge multiplets alone) were discussed from a field-theory perspective in \cite{Witten-sl2}. In general, one can also act on hypermultiplets, as discussed in \cite{DGG}.

Finally, decoupling one end of $M$ in the UV and one in the IR (on the Coulomb branch), we can obtain the 3d theory $T_K[M,\mb p,\Pi]$ for an ``RG wall'' \cite{DGV-hybrid}. It has the property that operators hitting the wall on the UV side are decomposed into a basis of IR operators on the other side, \cf\ \cite{Gaiotto-RG}. For supersymmetric line operators, such UV-IR maps have been discussed (\eg) in \cite{GMNIII, CN-line}, and RG walls give them a novel physical interpretation.
The 3d geometry representing an RG wall for $\CN=2^*$ theory is shown in Figure \ref{3d:fig:wallgeoms}c. (Note how two of these geometries can be glued along their outer boundaries to form the UV S-duality manifold of Figure \ref{3d:fig:wallgeoms}a.)


\section{3d theories, $SL(K)$ connections, and Chern-Simons}
\label{3d:sec:flat}

One of the most interesting geometric properties of a 3-manifold theory $T_K[M]$ is the relation between its vacua and flat $SL(K,\C)$ connections on $M$.
The other AGT-like correspondences between partition functions of $T_K[M]$ and Chern-Simons theory on $M$ in \eqref{3d:3d3d-intro} can be understood as quantizations of this basic semi-classical relation.
Strictly speaking, the relation to flat connections holds when $T_K[M]$ is compactified on a circle $S^1$ of finite radius. So let us do this, assuming that the full 6d geometry is now $M\times \R^2 \times S^1$.

The 6d (2,0) theory on a circle gives rise to 5d maximally supersymmetric Yang-Mills on $M\times \R^2$, with gauge group $SU(K)$,%
\footnote{\label{3d:foot:PSL}It is also possible to arrive at a theory where the center of $SU(K)$, or subgroups of the center, are not gauged. Then instead of getting a relation to $SL(K,\C)$ connections, we find a relation to $PSL(K,\C)$ connections, or similar. The details are subtle (see \cite{DGV-hybrid}), but the correct relation can ultimately be derived by examining the charges of fundamental line operators in $T_K[M]$.} %
and with a partial topological twist along $M$.
We may explicitly write down the 5d BPS equations. The partial twist transforms three real scalars in the gauge multiplet into an adjoint-valued 1-form $\varphi$ on $M$. The BPS equations on $M$ then take the form of ``Hitchin equations'' generalized to three dimensions:%
\footnote{The structure of Hitchin equations in two dimensions and their relation to 4d $\CN=2$ theory on a circle is reviewed in \cite{N}.}
\be [D_i,D_j] = 0\qquad (i,j=1,2,3)\,,\qquad \sum_{ij}g^{ij}[D_i,D_j^\dagger]=0\,, \label{3d:Hit3d} \ee
where $D_i$ is the gauge-covariant derivative with respect to a \emph{complexified} gauge field $A_i+i\varphi_i$, and $g_{ij}$ is a chosen background metric on $M$, \cf\ \cite{Wfiveknots, DGH, GW-Jones}. The equations are invariant under real $SU(K)$ gauge transformations. The set of solutions to \eqref{3d:Hit3d}, modulo real gauge transformations, is equivalent (up to a lower-dimensional subset) to the solutions of the equations $[D_i,D_j]=0$ alone, modulo \emph{complex} $SL(K,\C)$ gauge transformations. But this means that the solutions are complex $SL(K,\C)$ flat connections on $M$. Let us denote this moduli space as
\be \wt \CL_K(M) = \{ \text{flat $SL(K,\C)$ connections on $M$} \}\,. \label{3d:deftL} \ee
We expect it to correspond to the space of vacua of $T_K[M]$ on $S^1\times \R^2$.

In terms of branes, the M5 branes wrapping $M\times S^1\times \R^2$ become D4 branes wrapping $M\times\R^2$. The worldvolume theory of the D4's is 5d SYM. The three adjoint scalar fields that were promoted to a 1-form $\varphi$ are the translation modes of the D4's in the fibers of the cotangent bundle $T^*M$. In the infrared, one expects the stack of $K$ D4 branes to \emph{separate} in $T^*M$, becoming a single multiply-wrapped brane, and forming a spectral cover $\wt M$ of $M$. The pattern of separation then is encoded in the eigenvalues of the 1-form $\varphi$.

We might remark that starting from a flat complex connection $\CA$ and obtaining the spectral 1-form $\varphi$ is \emph{not} an easy task. To do so, one must find the right complex gauge transformation $h$ so that the transformed $\CA_h$ satisfies the real equation $g^{ij}[D_i,D_j^\dagger]=0$, in addition to the complex flatness equations. Then the imaginary part of this particular $\CA_h$ is $\varphi$. Therefore, $\varphi$ and the spectral cover it encodes depend on the choice of metric $g_{ij}$ for $M$ --- even though the notion of a flat complex connection does not.

It is also useful to observe that after splitting equations \eqref{3d:Hit3d} into real and imaginary parts they reduce to $F_A = dA+A^2 = \varphi^2$, along with $d_A\,\varphi = d_A*\varphi = 0$. The latter equations say that $\varphi$ is a covariantly harmonic one-form on $M$. The eigenvalues of $\varphi$ give rise (roughly) to a harmonic one-form on the spectral cover, which plays the role of a Seiberg-Witten form for~$T_K[M]$ \cite{CCV}.

If $M$ is compact and hyperbolic, the flat $SL(K,\C)$ connections on $M$ typically turn out to be rigid. We note, however, that mathematically it is still unknown precisely when rigidity holds.%
\footnote{One can attempt to use algebraic Mostow rigidity \cite{mostow-1973} to analyze the problem. This requires knowing that the representation $\rho:\pi_1(M)\to SL(K,\C)$ defined by the holonomies of a flat connection $\CA$ is a lattice. That is, $\rho(\pi_1(M))\subset SL(K,\C)$ is a discrete subgroup, with no accumulation points, such that $SL(K,\C)/\rho(\pi_1(M))$ has finite volume. This is true if $M$ is hyperbolic and $\CA$ is the flat connection related to the hyperbolic metric; but is unknown in general.} %
If the flat connections are indeed rigid, then $\wt \CL_K(M)$ consists of a discrete collection of points, and $T_K[M]$ will have isolated vacua (no moduli space) on $\R^2\times S^1$.

A more interesting situation arises when $M$ has asymptotic boundary $C$, so that $T_K[M]$ is a boundary condition for the 4d theory $T_K[C]$. Suppose that we move onto the Coulomb branch of $T_K[C]$. After compactification on a circle $S^1$, the theory $T_K[C]$ can be described in the IR as a 3d sigma model whose target is the moduli space of flat $SL(K,\C)$ connections on $C$ \cite{GMN}
\be \CP_K(C) = \{\text{flat $SL(K,\C)$ connections on $C$}\}\,.\label{3d:defP}\ee
This space arises physically from a standard 2d version of Hitchen's equations \eqref{3d:Hit3d}. It is actually a hyperk\"ahler space, as appropriate for 4d $\CN=2$ supersymmetry. However, we will only consider it in a single complex structure --- the complex structure associated to the 3d $\CN=2$ subalgebra that the boundary condition $T_K[M]$ preserves. Then, for us, $\CP_K[C]$ is simply a complex symplectic space. Its holomorphic symplectic form is given by the Atiyah-Bott formula
\be \Omega  = \int_C \Tr\big[ \delta \CA\wedge \delta \CA\big]\,, \label{3d:defO}\ee
where $\delta\CA$ is the deformation of a complex connection.
The holomorphic coordinates on $\CP_K[C]$ are eigenvalues or traces of $SL(K,\C)$ holonomies (or some more elementary cross-ratio coordinates, \emph{\`a la} \cite{FG-Teich}, from which holonomies can be constructed, see Section \ref{3d:sec:framed}). In $T_K[C]$ these coordinates are the vevs of supersymmetric line operators that wrap $S^1$ \cite{NRS, GMNIII}%
\footnote{See Section 2 of \emph{Hitchin systems in $\CN=2$ field theory} by A. Neitzke}.%

Now, the moduli space $\wt \CL_K(M)$ of flat connections on $M$ generically projects to a Lagrangian submanifold $\CL_K(M)\subset \CP_K(C)$, which parameterizes the flat connections on the boundary $C$ that extend to $M$:
\be \CL_K(M) = \{\text{flat $SL(K,\C)$ connections on $\pd M$ that extend to $M$}\}\,. \label{3d:defL} \ee
The expectation that this is Lagrangian follows from the fact that flatness equations are elliptic; at a basic level, only \emph{half} of the classical parameters on the boundary are needed to specify a flat connection in the bulk. Moreover, both $\CP_K(C)$ and $\CL_K(M)$ are algebraic. The equations that cut out $\CL_K(M)$ can be interpreted as Ward identities for line operators in $T_K[C]$ in the presence of the boundary condition $T_K[M]$.
In the effective 3d sigma-model to $\CP_K(C)$, $\CL_K(M)$ is quite literally a Lagrangian brane boundary condition~\cite{DGG}.

If we decouple the 4d bulk theory $T_K[C]$ to leave behind a 3d theory $T_K[M,\mb p]$ or $T_K[M,\Pi]$, the Lagrangian $\CL_K(M)$ acquires a more intrinsic interpretation. Let us consider the Seiberg-Witten description of $T_K[C]$ for simplicity. Then the choice of duality frame $\Pi$ needed for the decoupling maps precisely to a choice of \emph{polarization} for $\CP_K(C)$. This is a local splitting of coordinates into ``positions'' $x$ (corresponding to IR Wilson lines of $T_K[C]$) and ``momenta'' $p$ (corresponding to IR 't Hooft lines).

The decoupled theory $T_K[M,\Pi]$ has $U(1)^d$ flavor symmetry, where $2d=\dim_\C \CP_K(C)$. The positions $x$ are twisted masses%
\footnote{Explicitly, if we re-introduce the radius $\beta$ of the compactification circle, these dimensionless coordinates arise as  $x = \exp\big(\beta m_{3d}+i \oint_{S^1}A\big)$, where $A$ is the background gauge field for a 3d flavor symmetry, and $m_{3d}$ is its real mass. A factor of $\beta$ also enters \eqref{3d:LSUSY} to keep $\wt W$ dimensionless.} %
for each $U(1)$ symmetry, complexified by $U(1)$ Wilson lines around $S^1$. The momenta $p$ can be thought of as effective FI parameters for the flavor symmetries; or equivalently as the vevs of complexified moment map operators for each $U(1)$. The Lagrangian $\CL_K(M)$ then describes the subset of twisted masses and effective FI parameters that allow supersymmetric vacua to exist on $\R^2\times S^1$ --- it is the ``supersymmetric parameter space'' of $T_K[M,\Pi]$.

More concretely, by compactifying $T_K[M,\Pi]$ on a circle we obtain a 2-dimensional $\CN=(2,2)$ theory, whose IR behavior is governed by an effective twisted superpotential $\wt W$. After extremizing $\wt W$ with respect to dynamical fields, it retains a dependence on complexified masses $x$. The supersymmetric parameter space is then defined by \cite{DG-Sdual}%
\footnote{This Lagrangian and its quantization also plays a role in the study of surface operators in 4d $\CN=2$ theories, and their lifts to 3d defects in 5d theories --- see Section 2.4 of \cite{Gu}.}
\be \label{3d:LSUSY} \CL_K(M):\quad \exp \Big( x_i \pd_{x_i}\wt W(x)\Big) = p_i\,,\qquad i=1,...,d\,.\ee

The description of $\CL_K(M)$ and $\CP_K(\pd M)$ can be generalized to geometries $M$ that include codimension-two defects. It is necessary to impose boundary conditions for flat connections at the defects. These effectively increase the dimension of $\CP_K(\pd M)$, basically as if all defects had been regularized to small tubular pieces of boundary. This is natural, since defects enlarge the flavor symmetry group of $T_K[M]$. Mathematically, $\CP_K(\pd M)$ and $\CL_K(M)$ most accurately take the form of moduli spaces of ``framed'' flat connections, which we discuss in Section \ref{3d:sec:framed}.

\subsection{Quantization and 3d-3d relations}
\label{3d:sec:quant}

Having understood the fundamental relation between flat connections and the parameters/observables of $T_K[M]$, one can further deform the $\R^2\times S^1$ geometry to quantize the pair $\CL_K(M)\subset \CP_K(\pd M)$.
The basic idea is that adding angular momentum, so that $\R^2\simeq \C$ fibers over $S^1$ with twist $z\to q z$, leads to a non-commutative algebra of Wilson and 't Hooft line operators that satisfy $\hat p\hat x = q\, \hat x \hat p$ \cite{Ramified, GMNIII}, \cite[Section 3]{N}.
The algebraic equations for $\CL_K(M)$ are promoted to operators that annihilate partition functions of $T_K[M,\Pi]$ (or $T_K[M,\mb p]$), enforcing Ward identities in the twisted geometry.

The quantization of the pair $\CL_K(M)\subset \CP_K(\pd M)$ also has a natural interpretation on the ``geometric'' side of the 3d-3d correspondence. It is useful to recall that flat $SL(K,\C)$ connections on a 3-manifold are the classical solutions of quantum $SL(K,\C)$ Chern-Simons theory. The space $\CP_K(\pd M)$ is just the semi-classical phase space that Chern-Simons theory associates to a boundary of $M$, and its quantization produces the algebra of operators acting on a quantum Chern-Simons Hilbert space $\CH_K(\pd M)$ \cite{axelrod-witten, Witten-cx, gukov-2003}. Similarly, the Lagrangian $\CL_K(M)$ is just a semi-classical wavefunction, and its quantization produces a distinguished element of the operator algebra that annihilates the Chern-Simons wavefunction on $M$, an element of $\CH_K(\pd M)$.

One expects, therefore, that partition functions of $T_K[M,*]$ on spacetimes with angular momentum are equivalent to wavefunctions in complex Chern-Simons theory, leading to the correspondences of \eqref{3d:3d3d-intro}. A precise choice of spacetime is required to fully specify how the Chern-Simons Hilbert space should be quantized --- in particular to specify the level of the Chern-Simons theory. However, the structure of the quantum line-operator algebra (the algebra of operators in CS theory) remains essentially independent of this choice. Here are some options that have been studied:
\begin{itemize}

\item On spinning $\R^2\times_q S^1$ as above, the partition function of $T_K[M,\Pi]$ depends on a discrete choice $\alpha$ of boundary condition (basically a massive vacuum) at infinity on $\R^2$, in addition to $q$ and the complex masses $x$. Geometrically, $\alpha$ is a choice of flat connection on $M$ given boundary conditions $x$. The resulting partition functions $B_\alpha(x;q)$ \cite{Pasquetti-fact, BDP-blocks}, which count BPS states of $T_K[M,*]$, correspond to partition functions in analytically continued $SU(K)$ Chern-Simons theory on $M$, with exotic choices of integration contour labelled by $\alpha$, much as in \cite{gukov-2003, Wit-anal, Witten-path}.

\item The partition function of $T_K[M,*]$ on a spinning $S^2\!\times_q\! S^1$ geometry computes a supersymmetric index \cite{Kim-index, IY-index, KW-index}. It was conjectured in \cite{DGG-index} and derived in \cite{LY-S2} that the index corresponds to a wavefunction of $SL(K,\C)$ Chern-Simons theory at level $k=0$. This is not a trivial theory!
To be more precise, we must recall that complex Chern-Simons theory has two levels $(k,\sigma)$, one quantized and one continuous. Here only the quantized level is set to zero; the continuous $\sigma$ is related to the spin in the index geometry as $q \sim e^{2\pi /\sigma}$.

\item The partition function of $T_K[M,*]$ on an ellipsoid $S^3_b$, computed via methods of \cite{Kapustin-3dloc, HHL}%
\footnote{See also \emph{A review on SUSY gauge theories on $S^3$} by K. Hosomichi.}, %
was conjectured in \cite{Yamazaki-3d, DG-Sdual, DGG} to correspond to an $SL(K,\R)$-like Chern-Simons wavefunction. A careful supergravity calculation in \cite{CJ-sugra, CJ-S3} then derived a direct relation to $SL(K,\C)$ Chern-Simons theory at level $k=1$. The Hilbert spaces of these two Chern-Simons theories are very similar --- see Section \ref{3d:sec:AGT}.

\item It was conjectured in \cite{BDP-blocks} that the index and ellipsoid partition functions can both be written as sums of products of ``holomorphic blocks'' $B_\alpha(x;q)$, providing a direct relation between the three types of partition functions above. This is essentially holomorphic-antiholomorphic factorization in complex Chern-Simons theory, and involves a 3d analogue of topological/anti-topological fusion \cite{CV-tt*,CGV} for $T_K[M,*]$.

\item Extending the results of \cite{CJ-S3}, one expects that the partition function of $T_K[M,*]$ on a squashed lens space $L(k,1)_b$ (which can be computed via methods of \cite{BNY-Lens}) agrees with a wavefunction of $SL(K,\C)$ Chern-Simons theory at general level $k$.

\end{itemize}

The relation between $S^3_b$ partition functions of $T_2[M,*]$ and complex Chern-Simons theory provided some of the first concrete tests of 3d-3d duality.
For 3-manifolds with boundary, the relevant Chern-Simons partition functions could be computed using methods of \cite{hikami-2006, DGLZ, Dimofte-QRS} (and are now understood to capture $SL(2,\C)$ Chern-Simons at level $k=1$).
In the case of $S^2\times_qS^1$, however, techniques for computing the index of $T_K[M,*]$ led to a \emph{new} algorithm for computing $SL(K,\C)$ Chern-Simons wavefunctions at level $k=0$, which has since been formalized mathematically \cite{Gar-index, GHRS-index}. Repeating this exercise for squashed lens spaces should prove equally interesting.

\subsection{Connection to AGT}
\label{3d:sec:AGT}

As anticipated in the introduction, the fact that 3d theories $T_K[M,*]$ naturally define boundary conditions for 4d theories $T_K[\pd M]$ of class $\CS$ leads to a close interplay between the partition functions involved in 3d-3d and 2d-4d relations.

The basic physical idea is that if $X$ is a 4-manifold with boundary allowing supersymmetric compactification of $\CN=2$ theories, the partition function $\CZ_X\big(T_K[\pd M],\mb p\big)$ should depend on supersymmetric boundary conditions, and can be interpreted as a wavefunction in some Hilbert space $\CH_K[\pd M,\mb p]$. Here we write $\CZ_X\big(T_K[\pd M],\mb p\big)$ to emphasize that the way one prescribes boundary conditions may depend on a choice of weak-coupling duality frame for $T_K[\pd M]$, given (say) by a pants decomposition $\mb p$ for $\pd M$. For example, if $X=D^4_b$ is half of the squashed 4-sphere $S^4_b$ (equivalently, for computational purposes, to the omega-background $X=\R^4_\epsilon$), then $\CZ_X\big(T_2[\pd M],\mb p\big)$ is an instanton partition function of $T_2[\pd M]$. The instanton partition function depends on Coulomb moduli $a_i$ for each gauge group that is manifest in the duality frame $\mb p$. Via the AGT correspondence, it is natural to identify the instanton partition function with a wavefunction in the Hilbert space of Liouville conformal blocks $\CH_2[\pd M, \mb p]$.

Here we should emphasize a technical point. In this interpretation, $\CH_K[\pd M,\mb p]$ is not the (enormous) full physical Hilbert space of $T_K[\pd M]$ on $\pd X$.  Rather,  $\CH_K[\pd M,\mb p]$ is a ``BPS'' subsector of the full Hilbert space, whose elements are supersymmetric ground states of $T_K[M]$ on $\pd X$. The supersymmetric partition functions that we describe belong to this subsector, which has finite functional dimension.

Now if $M$ is any 3-manifold with boundary $\pd M$, then the partition function of $T_K[M,\mb p]$ on $\pd X$ should \emph{also} be a wavefunction in the Hilbert space $\CH_K[\pd M,\mb p]$. In order to calculate the partition function of $T_K[\pd M]$ on $X$, coupled to the theory $T_K[\pd M,\mb p]$ on $\pd X$, we simply take an inner product
\be \label{3d:IP}
  \big\langle  \CZ_X\big(T_K[\pd M],\mb p\big)\,\big|\,
  \CZ_{\pd X}\big(T_K[M,\mb p]\big)\, \big\rangle\,.
\ee
For example, if $X=D^4_b$, then $\pd X=S^3_b$, and $\CZ_{\pd X}\big(T_2[M,\mb p]\big)$ is simply the ellipsoid partition function of the 3d theory $T_2[M,\mb p]$. Note that the 3d theory $T_2[M,\mb p]$ has flavor symmetries with complexified twisted masses $a_i$ for every gauge symmetry of the bulk theory $T_2[\pd M]$ (in duality frame $\mb p$); thus both the right and left sides of \eqref{3d:IP} depend on the same parameters $a_i$, and taking an inner product just means integrating them out with the right measure.

By using the doubling trick of Figure \ref{3d:fig:domSW}, these constructions can easily be extended to domain walls. For example, one might insert an S-duality domain wall carrying theory $T_2[M;\mb p,\mb p']$ on the equator $S^3_b \subset S^4_b$. Here $\pd M=\ol C\sqcup C$ for some surface $C$, so the ellipsoid partition function belongs to a product of Hilbert spaces $\CZ_{S^3_b}\big(T_2[M;\mb p,\mb p']\big) \in \CH_2[C]^*\otimes \CH_2[C]$. The partition function on the whole $S^4_b$ with the domain wall becomes
\be \big\langle \CZ_{D^4_b}\big(T_2[C],\mb p\big)\,\big|\, \CZ_{S^3_b}\big(T_2[M,\mb p,\mb p']\big)\, \big|\, \CZ_{D^4_b}\big(T_2[C],\mb p'\big)\, \big\rangle\,.
\ee
Such configurations with S-duality domain walls in $S^4_b$ have been studied at length, \eg\ in \cite{HLP-wall, Yamazaki-3d, DG-Sdual, TV-6j} (see \cite{H}).
The 3d partition function $\CZ_{S^3_b}\big(T_2[M,\mb p,\mb p']\big)$ can be identified with a Moore-Seiberg kernel in Liouville theory --- it acts naturally on $\CH_2[C]$, changing the basis from one labelled by $\mb p$ to one labelled by $\mb p'$. In this case, the Lagrangian $\CL_2(M)$ and its quantization describes the transformation of line operators from one side of the wall to the other. An analogous setup involving domain walls on the equator of the index geometry $S^2\times_qS^1\subset S^3\times_q S^1$ was considered in \cite{DGG-index, Gang-wall, Gang-index}.

We remark that while physically it is clear that all wavefunctions appearing in formulas such as \eqref{3d:IP} must belong to the same Hilbert space --- namely, the space describing basic supersymmetric boundary conditions on $\pd X$ --- this is sometimes a little less clear on the ``geometric'' side of the 3d-3d and 2d-4d correspondences. There remain a few interesting details to be worked out here. For example, the 2d-4d correspondence says that $T_2[\pd M]$ belongs to a space of Liouville conformal blocks on $\pd M$, while the 3d-3d correspondence says that $T_2[M,*]$ belongs to the Hilbert space of $SL(2,\C)$ Chern-Simons theory, at level $k=1$, on $\pd M$. These are not obviously equivalent. A promising observation is that the Liouville Hilbert space is a boundary Hilbert space for $SL(2,\R)$ Chern-Simons \cite{Verlinde-TeichLiouv}%
\footnote{Quantization of $SL(2,\R)$ flat connections on a surface is reviewed in this volume in \emph{Supersymmetric gauge theories, quantization of $\CM_{\rm flat}$, and conformal field theory} by J. Teschner.}. %
In turn, the quantization of a model phase space $(\R)^2$ in $SL(2,\R)$ theory yields $\CH = L^2(\R)$; while the quantization of a model $(\C^*)^2$ in $SL(2,\C)$ theory at level $k$ yields $\CH=L^2(\R)\otimes V_k$, where $\dim V_k=k$ \cite{DGG-index};  these model descriptions agree when $k=1$.


\section{Top-down construction}
\label{3d:sec:topdown}

Currently there exist two closely related approaches for producing 3d $\CN=2$ Lagrangian gauge theories that flow in the IR to 3-manifold theories $T_K[M]$. Both approaches lead to abelian Chern-Simons-matter theories of class $\CR$, whose superpotentials may contain nonperturbative monopole operators. Going in reverse chronological order, we will first introduce the more intuitive ``top-down'' construction of \cite{CCV} here, and then discuss the more concrete but also more technical ``bottom-up'' construction of \cite{DGG, DGG-Kdec} in Section \ref{3d:sec:bottomup}.

It is important to keep in mind that many different UV Lagrangian theories can have the same IR fixed point $T_K[M]$. We will say that such UV theories are ``mirror symmetric,'' after the first dualities of this type found in \cite{IS, AHISS, dBHOY, dBHO}. The phenomenon is entirely analogous to Seiberg duality for 4d $\CN=1$ theories (and sometimes even arises from reducing 4d dualities \cite{ARSW-dualities}). For now we note that the abelian Lagrangians described here could easily have non-abelian mirrors.

The basic idea of \cite{CCV} is to derive the BPS particle content and interactions for a UV description of $T_K[M]$ from the geometry of a $K$-fold spectral cover $\wt M$ of $M$, and then to use them (optimistically) to reconstruct an entire 3d Lagrangian. For example, in M-theory, the $K$ coincident M5 branes wrapping $M$ are expected to deform at low energy%
\footnote{Here we mean low energy from the point of view of M-theory dynamics, which is still UV for 3d field theories on $\R^3$. See related comments below about being able to choose arbitrary metric for $M$.} %
in the fiber directions of $T^*M$, recombining into a single brane that wraps the cover $\wt M$. The BPS states and their interactions then arise from M2 branes that end on this M5.

We can understand the appearance of a spectral cover $\wt M$, governed by a multi-valued harmonic one-form $\lambda$ on $M$ (or a single-valued harmonic one-form $\lambda$ on $\wt M$), directly in $M$ theory. In order to preserve supersymmetry, an M5 brane must wrap a special Lagrangian 3-cycle in $T^*M$. The zero-section $M$ is one such cycle, but it can be deformed. Small deformations preserving the special Lagrangian condition are precisely parametrized by real harmonic 1-forms $\lambda$ on $M$. We should emphasize again that $\lambda$ depends on a choice of metric for $M$, which is entirely up to us --- we are \emph{not} working in the ultra low-energy limit where only the hyperbolic metric is relevant.

Alternatively, we could obtain the spectral cover in field theory by starting with the nonabelian Hitchin-like construction of Section \ref{3d:sec:flat}, and sending the compactification radius to infinity. This radius $\beta$ implicitly entered the definition of the complexified connection $A+i\beta\varphi$ in \eqref{3d:Hit3d}; as $\beta\to \infty$, a rescaled Higgs field $\varphi$ survives. So long as the three components of $\varphi$ are simultaneously diagonalizable, we saw that their eigenvalues define a multi-valued harmonic 1-form. A more direct 6d construction, along the lines of \cite{Witten-M, Gaiotto-dualities}, would extract $\lambda$ from certain operators of the 6d $(2,0)$ theory.

From $\wt M$ and $\lambda$, one can attempt to read off the content of a UV Lagrangian description of $T_K[M]$, which we'll call $\wt T_K[M]$. First, the integral of $\lambda$ around any 1-cycle $\gamma \subset \wt M$ produces a real scalar $\sigma$ in a 3d $\CN=2$ vector multiplet. The integral of the (abelian) M5-brane $B$-field on the same cycle leads to the actual 3d abelian gauge field $A_\mu$, the superpartner of $\sigma$.
Thus, to a first approximation, the number of gauge multiplets in $\wt T_K[M]$ is the first Betti number $b_1(M)$. In fact, if there is any torsion in $H_1(\wt M,\Z)$, it indicates the presence of additional gauge multiplets that are killed (dynamically) by nonzero Chern-Simons terms. The full claim is that if
\be H_1(\wt M,\Z) \simeq \Z\langle \gamma_1,...,\gamma_d\rangle\big/\big(\Sigma_j k_{ij}\gamma_j=0\big)\,,
\ee
then $\wt T_K[M]$ has $d$ abelian gauge multiplets coupled with a Chern-Simons level matrix $k_{ij}$.

If $M$ has defects, they lift to defects in the spectral cover $\wt M$. Then, much as in the setting of compactification on 2d surfaces with punctures, the non-trivial 1-cycles in $\wt M$ that link the defects give rise to non-dynamical gauge fields and flavor symmetries in $\wt T_K[M]$. Note that defects impose boundary conditions on $\lambda$ that forbid a trivial solution $\lambda\equiv 0$.

Similarly, if $M$ has an asymptotic boundary of the form $C\times \R_+$, the spectral cover $\wt M$ will have asymptotic regions of the form $\Sigma\times \R_+$, where $\Sigma$ is a $K$-fold cover of $C$. It is the Seiberg-Witten curve for the 4d theory $T_K[C]$. If we pass to a weak-coupling limit $\Pi$ of $T_K[C]$ to obtain a pure 3d theory $T_K[M,\Pi]$, half of the cycles in the Seiberg-Witten curve will get pinched off. The remaining cycles contribute to $H_1(\wt M)$, and lead to non-dynamical $U(1)$ gauge multiplets in $\wt T_K[M,\Pi]$, corresponding to the expected $U(1)$ flavor symmetries.

\begin{figure}[htb]
\centering
\includegraphics[width=5.5in]{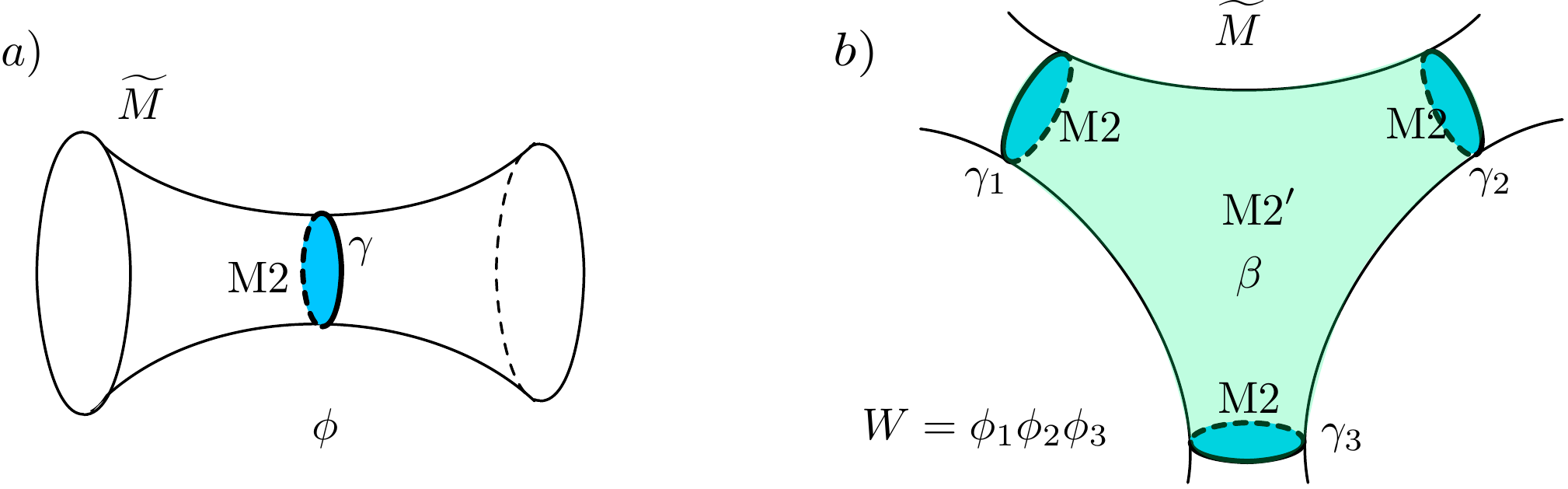}
\caption{\small Producing BPS chirals and superpotentials from M2 branes wrapped on $\wt M$.}
\label{3d:fig:M52}
\end{figure}

Most interestingly, M2 branes ending on the M5 wrapping the spectral cover lead to BPS particles and superpotential interactions in $T_K[M]$. The basic case is a non-contractible cycle $\gamma\subset \wt M$ of \emph{minimum volume} that bounds a disc $D_\gamma\subset T^*M$. An M2 brane wrapping $D\times \R\subset T^*M\times \R^3$ gives rise to a BPS particle of charge $\gamma$, hence a chiral multiplet $\phi$ in $\wt T_K[M]$ (Figure \ref{3d:fig:M52}a). If the M2 brane instead ends on a 2-cycle $\beta\subset \wt M$ (filling in a ball in $T^*M$), then it looks like an instanton in $\R^3$, which can generate a superpotential involving a monopole operator. It is the monopole for the gauge field associated to the 1-cycle $\gamma$ dual to $\beta$. Finally, suppose we have a collection of M2 branes wrapping some discs $D_i$, with $\pd D_i=\gamma_i$, giving rise to chirals $\phi_i$. Then an additional M2' brane might wrap a ball in $T^*M$ whose boundary is a union of the discs $D_i$ and an open 2-cycle in $\wt M$ connecting their boundaries $\gamma_i$ (Figure \ref{3d:fig:M52}b). This latter M2' brane also looks like an instanton in $\R^3$, and generates a superpotential interaction among the chirals, $W = \prod_i \phi_i$.

Altogether, the vector multiplets and their Chern-Simons interactions, and the chiral multiplets and their superpotential interactions, all obtained geometrically from $\wt M$, could specify the abelian Chern-Simons matter theory $\wt T_K[M]$ (or $\wt T_K[M,\Pi]$, etc.). Unfortunately, the prescription can be extremely difficult to implement in general. The problem is that, given an arbitrary background metric on $M$, one cannot easily solve for the harmonic form $\lambda$ and the minimum-volume cycles on $\wt M$.

One way to circumvent this problem is to deform the metric on $M$ so that the cover $\wt M$ becomes ``especially nice,'' making it easy to read off the particle content of $\wt T_K[M]$. We will explain this further in the next sections. Often there are multiple ``especially nice'' limits, which lead to different mirror-symmetric theories $\wt T_K[M]$.

\subsection{Seiberg-Witten domain walls}
\label{3d:sec:SWwall}


A basic scenario that can allow a simple description of the spectral cover $\wt M$ is for $M$ representing a Seiberg-Witten domain wall, as discussed in Section \ref{3d:sec:duality}. Such manifolds were the focus of study in \cite{CCV, Cordova-tangles}.

We take $M=\R\times C$, where $C$ is a punctured surface. The punctures become defects running the entire length of $M$. At the two asymptotic ends of $M$, we consider the theory $T_K[C]$ on its Coulomb branch. Globally, we picture the spectral cover $\wt M$ as a fibration over the infinite direction $\R$, whose fiber over a point $x^3\in \R$ is a Seiberg-Witten curve $\Sigma_{x^3}$ for $T_K[C]$. The Seiberg-Witten curve comes with a holomorphic Seiberg-Witten differential $\lambda^{SW}(x^3)$. As $x^3$ varies from $-\infty$ to $\infty$, we want to smoothly vary the UV gauge couplings (\ie\ the metric on $C$), as well as mass parameters coming from the defects and Coulomb moduli in such a way that the theory $T_K[C]$ decouples at $x=\pm \infty$ according to some chosen polarizations $\Pi,\Pi'$.

In order to preserve 3-dimensional $\CN=2$ supersymmetry, the variation we choose cannot be completely arbitrary. Geometrically, we need the real part of the varying Seiberg-Witten differential $\lambda^{SW}(x^3)$ to form two of the three components of a harmonic 1-form $\lambda$ on $\wt M$. Alternatively, in field-theory terms, we recall that the 3d $\CN=2$ central charges are the real parts%
\footnote{More generally, we have $Z_{3d}=\Re[\zeta^{-1}Z_{4d}]$, where the phase $\zeta$ characterizes the $4d\to 3d$ supersymmetry breaking. The 4d R-symmetry group $SU(2)_R\times U(1)_r$ is broken to $U(1)_R$ (a Cartan of $SU(2)_R$), and this $\zeta$ is rotated by the broken $U(1)_r$. This same phase also happens to select the complex structure that one should use for the hyperk\"ahler moduli spaces of flat connections \cite{GMN, GMNII}, as discussed in Section \ref{3d:sec:flat}.} %
of 4d $\CN=2$ central charges (just as the scalar in a 3d gauge multiplet is the real part of the scalar in a 4d gauge multiplet). A necessary condition for unbroken 3d SUSY is
\be \label{3d:Janus} \pd_3 \Re[ a(x_3)] = \pd_3 \Re[ a_D(x_3)] = \pd_3 \Re[m(x_3)]= 0\,, \ee
\ie\ the real parts of all 4d central charges, coming from periods of $\lambda^{SW}$, are fixed.
A 4d theory $T_K[C]$ whose parameters vary%
\footnote{Similar half-BPS configurations in 3d $\CN=2$ theories were discussed in \cite{GGP-walls}.} %
in the $x^3$ direction subject to \eqref{3d:Janus} can be called a generalized \emph{Janus configuration}, \cf\ \cite{GW-Janus}.
The condition \eqref{3d:Janus} ensures that $\pd_3\Re\,\lambda^{SW}$ is an exact 2-form on $\Sigma$, \ie\ $\pd_3\Re\,\lambda^{SW}=d_\Sigma f$, where $d_\Sigma$ is the exterior derivative along $\Sigma$. Then $\Re\,\lambda^{SW}-f\,dx^3$ is a closed real 1-form on $\wt M$, which can be further corrected%
\footnote{The correction requires solving the potential problem $\nabla^2\sigma = \pd_3f$. Then $\lambda=\Re\,\lambda^{SW}-f\,dx^3+d\sigma$.} %
to produce the harmonic 1-form~$\lambda$.

The fundamental example of a Seiberg-Witten domain wall involves the Seiberg-Witten curve
\be \Sigma_\Delta:\quad z^2 = -w^2+m\,,\qquad \lambda^{SW} = z\,dw\,, \label{3d:A1} \ee
where $m$ is a complex mass parameter. Note that the curve is a double cover of the complex $w$-plane, which we identify as $C$, with branch points at $w=\pm \sqrt{m}$, and that the only nontrivial period comes from the cycle $\gamma$ connecting the branch points:
\be \frac{1}{\pi}\oint_\gamma \lambda^{SW} = \frac{2}{\pi}\int_{-\sqrt{m}}^{\sqrt{m}}\lambda^{SW} =  m\,. \ee
Indeed, the Seiberg-Witten theory corresponding to the curve \eqref{3d:A1} has a single BPS hypermultiplet of central charge $m$ (and mass $|m|$). More generally, the curve \eqref{3d:A1} can also be thought of as a local model for any Seiberg-Witten fibration $\Sigma\to C$ where two branch points are coming close together.

To build a domain wall from \eqref{3d:A1}, we vary the imaginary part of $m$ while keeping the real part fixed, say $m= m_0 + i x_3$. The two branch points of $\sigma\to C$ sweep out branch lines of a 3d fibration $\wt M\to M$. As $x_3\to \pm \infty$, the branch lines move very far apart, the mass $|m|$ of the 4d BPS state grows infinitely, and the 4d theory $T_K[C]$ decouples. At $x_3=0$, the branch lines are minimally separated, and an M2 brane wrapping the cycle $\gamma$ between them produces a ``trapped'' 3d BPS chiral $\phi$. Its 3d real mass is $m_0$. We find that $T_K[M,\Pi,\Pi'] =: T_\Delta$ (which will eventually be called the ``tetrahedron theory'') contains a single free chiral transforming under the $U(1)$ flavor symmetry coming from the cycle $\gamma$. If we want a true SCFT, we should set $m_0=0$; otherwise the 3d theory is mass-deformed.

In field-theory terms, the full domain wall $T_K[M]$, can be understood roughly as follows.  Let us denote by $T_K[C^-]$ and $T_K[C^+]$ the 4d Seiberg-Witten theories on the left and right half-spaces $\R^3\times \R_{\pm}$. Each of these theories has a BPS hypermultiplet $\Phi^-$ and $\Phi^+$, which we rewrite as a pair of 3d $\CN=2$ chirals $(X^-,Y^-)$ and $(X^+,Y^+)$. Here $X$ and $Y$ have opposite flavor charge. On both the left and the right, we give $X^\pm$ Dirichlet boundary conditions and $Y^\pm$ Neumann boundary conditions. Then, at $x_3=0$, we couple the (free) boundary values of $Y^\pm$ to our 3d chiral $\phi$ via a superpotential \cite{DGG}
\be W = Y^-\phi - \phi Y^+\,\big|_{x_3=0}. \label{3d:Wdelta} \ee
These couplings modify the Dirichlet b.c. for the $X$'s to $X^-|_{x^3=0}=\phi = X^+|_{x^3=0}$, via a mechanism studied in \cite{GW-boundary, dWFO}.

In the far infrared, we can simply use \eqref{3d:Wdelta} to integrate out $\phi$, obtaining $Y^+=Y^-$ and $X^+=X^-$. Thus we recover a single 4d theory $T_K[C]$ on all of $\R^4$. This is not unexpected: in the deep IR, all Seiberg-Witten ``duality'' walls are basically trivial! However, if we first send $\Im\, m\to \infty$ on the left and right sides of the wall to freeze out the 4d hypers, we are left with the decoupled 3d theory $T_\Delta$ containing a nontrivial chiral $\phi$.

Note that the choices $\Pi$ and $\Pi'$ that we made to decouple the two sides in this example had nothing to do with dynamical electric/magnetic gauge fields. They simply selected which halves of the hypers $(X^\pm,Y^\pm)$ got Neumann vs. Dirichlet boundary conditions.%
\footnote{It may seem like $\Pi=\Pi'$ in this example. This is not the case, due to the relative orientation on the two halves. The setup corresponding to $\Pi=\Pi'$ involves $X$ getting Dirichlet b.c. on one side and $Y$ getting Dirichlet b.c. on the other, with the remaining (Neumann) halves coupled directly by a superpotential $W=Y^-X^+$ at $x^3=0$. This flows immediately to $T_K[C]$ on all of $\R^4$.} %
More generally, one may augment couplings to 3d chirals as in \eqref{3d:Wdelta} with true changes of polarization, which are implemented by pure 3d $\CN=2$ Chern-Simons theories living on the domain wall \cite{Witten-sl2} (see also Section \ref{3d:sec:TD}).

\section{Bottom-up construction: symplectic gluing}
\label{3d:sec:bottomup}

In the last section, we mentioned that a judicious choice of metric on $M$ can lead to an especially simple spectral cover $\wt M$, so that the full abelian Chern-Simons Lagrangian of a theory $\wt T_K[M]$ can be read off. What we had in mind was a cover branched along a set of lines, so that the branch lines are well separated almost everywhere. In a few isolated regions, the branch lines pass close by one another, and each such region might be modeled on the example \eqref{3d:A1} of Section \ref{3d:sec:SWwall}. Graphically, each region of closest-approach may be represented as a tetrahedron $\Delta$ in a 3d triangulation of $M$. Then we can attempt to associate a canonical ``tetrahedron theory'' $T_\Delta$ to each tetrahedron --- basically the theory of a free 3d chiral multiplet --- and then to glue them together properly. This is what was done in \cite{DGG} for $K=2$, and generalized to arbitrary $K\geq 3$ in \cite{DGG-Kdec}.

The idea of \cite{DGG} was to develop a complete, consistent set of gluing rules for tetrahedron theories, working from the ground up. Physically, the gluing rules amount to introducing superpotential couplings for internal edges in a triangulated manifold, and possibly gauging $U(1)$ flavor symmetries. The rules are very precise, and make many properties of $T_K[M]$ manifest --- such as the presence of various marginal and relevant operators, and the existence of an unbroken $U(1)_R$ symmetry in the infrared.  On the other hand, one always obtains abelian Chern-Simons matter Lagrangians with abelian flavor symmetries, and it can be quite nontrivial to see that some of the flavor symmetries have expected nonabelian enhancements, \eg\ to $SU(K)$. More seriously, as mentioned in the introduction, the theories obtained from triangulations sometimes capture only a sub-sector of the full $T_K[M]$; we will explain why in Section \ref{3d:sec:sub}.

Geometrically, the approach of \cite{DGG} mimics a construction of classical and quantum flat $SL(K)$ connections on 3-manifolds via ``symplectic gluing.'' The method of symplectic gluing for quantized connections on triangulated manifolds was developed in \cite{Dimofte-QRS}, generalizing classical observations of Neumann and Zagier \cite{NZ} and Thurston \cite{thurston-1980} in hyperbolic geometry. The basic idea, going back to work of Atiyah and A. Weinstein, is that when gluing $M=M_1\cup_\Sigma M_2$ along some boundary $\Sigma$, the standard notion of ``taking an inner product of wavefunctions in boundary Hilbert spaces'' can be replaced by a formally equivalent procedure of quantum symplectic reduction. The latter procedure is easy to implement even when only partial pieces of boundary are glued.

Since the gluing rules for theories $T_K[M]$ are built to match the gluing of quantum connections, many of the relations between sphere partition functions of $T_K[M]$ and Chern-Simons wavefunctions on $M$ that were summarized in Section \ref{3d:sec:flat} can be proven combinatorially. More interestingly, one realizes that for a manifold $M$ with boundary, the theory $T_K[M,\Pi]$ should \emph{itself} be viewed as a sort of wavefunction --- with its flavor symmetries playing the role of ``position variables'' that the wavefunction depends on.

We proceed to summarize some of the results of \cite{Dimofte-QRS, DGG, DGG-Kdec}, starting with symplectic gluing in geometry and then extending the gluing to 3d gauge theory.

\subsection{Framed 3-manifolds and framed flat connections} 
\label{3d:sec:framed}

It is useful to introduce a topological class of \emph{framed 3-manifolds} \cite{DGG-Kdec, DGV-hybrid}, which represent the 3-manifolds with asymptotic boundaries and networks of defects from Section \ref{3d:sec:6d} that were used to compactify the 6d $(2,0)$ theory. A framed 3-manifold%
\footnote{Such manifolds were called ``admissible'' in \cite{DGG-Kdec}.} %
is a 3-manifold $M$ with non-empty boundary $\pd M$, along with a separation of $\pd M$ into ``big'' and ``small'' pieces:
\begin{itemize}
\item The big boundary consists of surfaces $C$ of arbitrary genus $g$ and $h\geq 1$ holes, such that $-\chi(C)=2g-2+h>0$. (In particular, these surfaces admit 2d hyperbolic metrics.)
\item The small boundary consists of discs, annuli, or tori. The $S^1$ boundaries of small discs and annuli connect to the holes on the big boundary.
\end{itemize}
Each of the big boundaries $C$ is meant to represent an asymptotic boundary of a compactification manifold --- or rather an asymptotic boundary that has been ``cut off'' to isolate a 3d theory. Each of the small boundaries represents a codimension-two defect that has been regularized to a long, thin tube.

\begin{figure}[htb]
\centering
\includegraphics[width=6in]{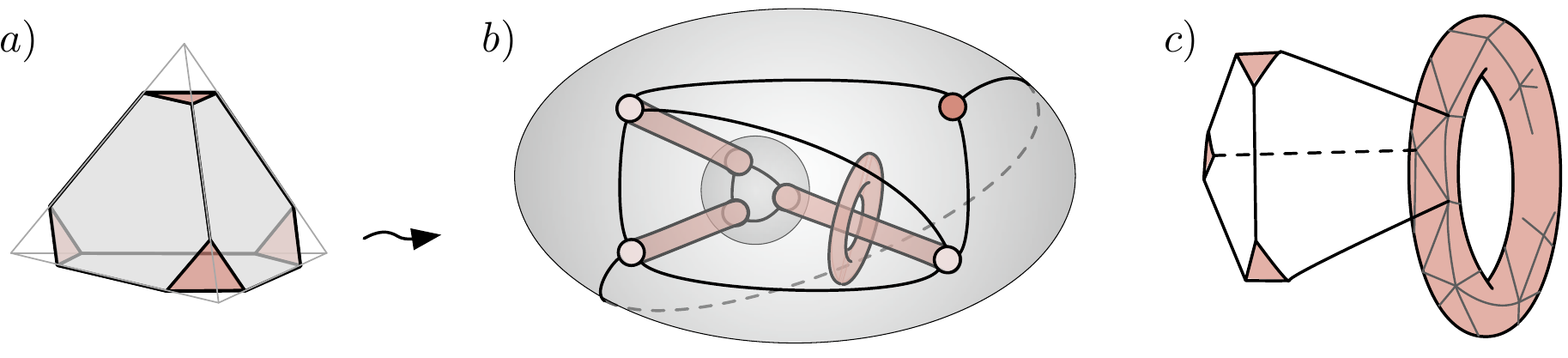}
\caption{\small Truncated tetrahedra (a), which can be glued together to form a framed 3-manifold $M$ (b). The small vertex-triangles of tetrahedra tile the small tubular boundaries of $M$ (c).}
\label{3d:fig:admM}
\end{figure}

An oriented framed 3-manifold can be glued together from oriented, truncated tetrahedra (Figure \ref{3d:fig:admM}), which are themselves framed 3-manifolds. The big boundary of a tetrahedron is a 4-holed sphere, tiled by four big hexagons. The small boundary consists topologically of four small discs, the triangular vertex neighborhoods. In order to form any more complicated framed 3-manifolds, the big hexagons on tetrahedron faces are glued together in pairs --- so some parts of the big boundary may remain unglued --- while the small boundary is \emph{never} glued.

Notice that a 3d triangulation of a framed 3-manifold induces a 2d ``ideal triangulation'' of its big boundary, \ie\ a triangulation where all edges begin and end at the holes/punctures. Having fixed the big-boundary triangulation, all possible 3d triangulations of the interior are related by performing sequences of 2--3 moves, shown below in Figure \ref{3d:fig:23}.

Geometrically, on a framed 3-manifold $M$ we can study \emph{framed flat connections}. This is a precise mathematical object that ultimately reproduces (an algebraically open subset of) the correct supersymmetric parameter space of a theory $T_K[M,\Pi]$ on a circle, refining \eqref{3d:deftL}--\eqref{3d:defL}. (Framed flat connections in two dimensions played a prominent role in \cite{FG-Teich, GMN-snakes}.)

A framed flat $PSL(K,\C)$ connection on $M$ is a standard flat $PSL(K,\C)$ connection together with a choice of invariant flag on each small boundary component. It might be useful to recall that a flag is a set of nested subspaces
\be  \{0\}\subset F_1\subset \cdots\subset F_K = \C^K\,,\qquad \dim F_K=K\,. \ee
For example, a flag in $\C^2$ is just a complex line in $\C^2$, \emph{a.k.a.} a point in $\cp^1$. What we require for the framing of a flat connection is a choice of flat section of an associated flag bundle on $\pd M$ that's invariant under the $PSL(K,\C)$ holonomy around each small boundary. Then we set
\begin{align} \label{3d:PL}
 \CP_K(\pd M) &= \{\text{framed flat $PSL(K,\C)$ connections on $\pd M\bs$(all small discs)}\}\,, \\
 \CL_K(M) &= \{\text{connections in $\CP_K(\pd M)$ that extend to framed flat connections on $M$}\}\,.\notag
\end{align}
As discussed in Footnote \ref{3d:foot:PSL}, one sometimes needs to lift these spaces to $SL(K)$ rather than $PSL(K)$, depending on the precise theory of interest. Here we will use $PSL(K)$ for concreteness.

The choice of framing for a flat connection is usually unique, or almost so. For example, a $PSL(K)$ holonomy matrix with distinct eigenvalues has a unique set of $K$ eigenvectors. Choosing an ordering of the eigenvectors, one can then construct an invariant flag. On the other hand, if eigenvalues coincide there may be a continuous choice of invariant flag. This choice resolves singularities in the naive moduli spaces $\CP_K(M)$, $\CL_K(M)$. An analogous physical resolution of moduli spaces is well known to exist in the presence of defects on surfaces, \cf\ \cite{Ramified, Wit-anal, GMNII}.

The fundamental example of a framed pair \eqref{3d:PL} is for a truncated tetrahedron $\Delta$, with $K=2$. On the boundary $\pd \Delta$, viewed as a sphere with four holes, we consider framed flat connections with unipotent holonomy around the holes. (It is necessary to ask for unipotent holonomy, \ie\ unit eigenvalues, in order for flat connections to potentially extend to the interior.) At each hole, we choose a complex line in $\C^2$ that's an eigenline of the holonomy there. If the holonomy is parabolic, of the form $\large \left(\begin{smallmatrix} 1 & a \\ 0 & 1 \end{smallmatrix}\right)$ with $a\neq 0$, the eigenline is unique. On the other hand, if the holonomy becomes trivial $\large\left(\begin{smallmatrix} 1 & 0 \\ 0 & 1\end{smallmatrix}\right)$, the eigenline is completely undetermined. This extra choice in the latter scenario blows up a singularity in the unframed moduli space.

\begin{figure}[htb]
\centering
\includegraphics[width=5.3in]{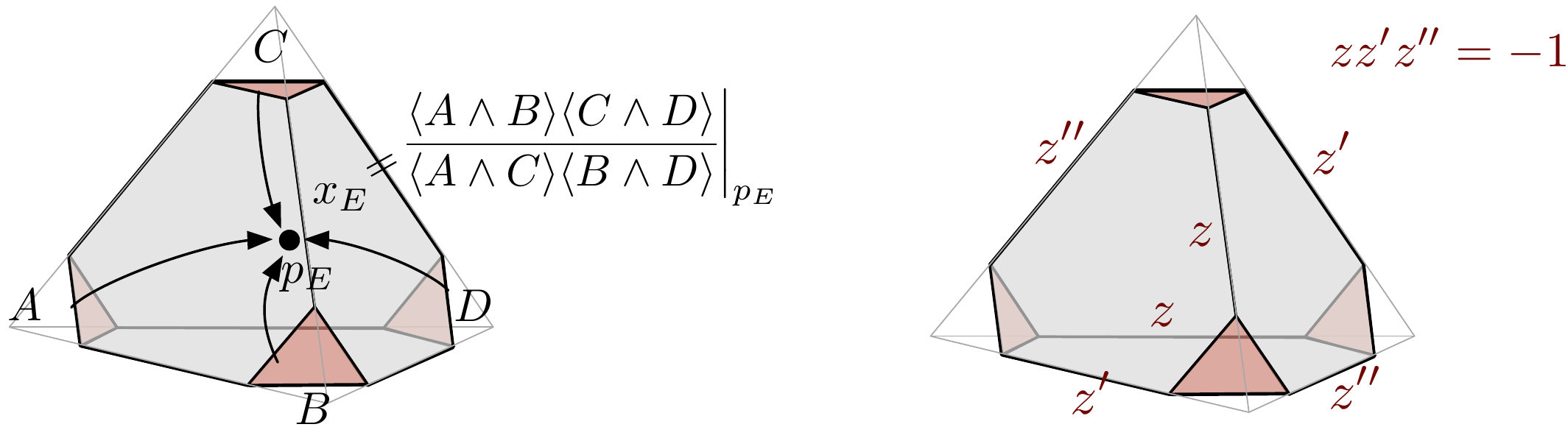}
\caption{\small Defining six edge-coordinates for a tetrahedron by parallel-transporting lines $A,B,C,D$ to common points $p_E$, then taking cross-ratios.}
\label{3d:fig:tetz}
\end{figure}

We can parametrize a generic framed flat $PSL(2)$ connection on $\pd \Delta$ with ``cross-ratio coordinates'' of Fock and Goncharov \cite{FG-Teich}, as follows.%
\footnote{These coordinates generalize Thurston's classic shear coordinates in Teichm\"uller theory, later studied by Penner, Fock, and others.} %
Every edge $E$ in the natural triangulation of $\pd \Delta$ is contained in a unique (truncated) quadrilateral. We parallel-transport the eigenlines at the four vertices of this quadrilateral to any common point $p_E$ inside the quadrilateral, and take their cross-ratio%
\footnote{Recall that lines in $\C^2$ are just points in $\cp^1$, so an $SL(2)$-invariant cross-ratio can be formed.} %
to define a coordinate $x_E$. The product of these cross-ratio coordinates around any tetrahedron vertex is $-1$ (due to the unipotent holonomy), which also implies that coordinates on opposite edges are equal. Relabeling the edge-coordinates $z,z',z''$ as on the left of Figure \ref{3d:fig:tetz}, we find that
\be \CP_2(\pd \Delta) \approx \big\{z,z',z'' \in \C^*\,\big|\,zz'z''=-1\big\} =: \CP_{\pd \Delta}\,, \label{3d:PD}\ee
with expected complex dimension 2.
The complex symplectic structure on $\CP_{\pd \Delta}$ induces Poisson brackets $\{\log z,\log z'\}=\{\log z',\log z''\}=\{\log z'',\log z\}=1$.

Similarly, we may consider framed flat connections in the bulk of $\Delta$. But now, since $\Delta$ is contractible, any flat connection is gauge-equivalent to a trivial one. Nevertheless, the choice of four eigenlines at the vertices (modulo the overall action of $PSL(2)$) remains, and is parametrized by the Lagrangian submanifold
\be \CL_\Delta = \{z''+z^{-1}-1=0\} \subset \CP_{\pd\Delta}\,. \label{3d:LD} \ee
The relation $z''+z^{-1}-1=0$ (which could equivalently be written as $z+z'{}^{-1}-1=0$ or $z'+z''{}^{-1}-1=0$) is simply a standard Pl\"ucker relation among the cross-ratio coordinates, reflecting the fact that after the tetrahedron is filled in we may parallel-transport all eigenlines to a common point in the interior of $\Delta$ and simultaneously calculate all cross-ratios there.

For a general framed 3-manifold $M$, we may choose a 2d triangulation of the big boundary and again construct cross-ratio coordinates $x_E$ there. Their Poisson bracket is such that
\be \{\log x_E,\log x_{E'}\} = \text{oriented $\#$ of faces shared by $E,E'$}\,.\ee
These are supplemented by holonomy eigenvalues around A- and B-cycles of small torus boundaries, and by a combination of holonomy eigenvalues and canonically conjugate ``twist'' coordinates for each small annulus, altogether forming a system of coordinates for an algebraically open patch of $\CP_2(\pd M)$ that's isomorphic to a complex torus $(\C^*)^{2d}$. The fundamental result is that if $M$ is cut into $N$ truncated tetrahedra (in any manner that's consistent with the chosen boundary triangulation) then this patch of $\CP_2(\pd M)$ is a symplectic quotient
\be \CP_2(\pd M)= \Big(\prod_{i=1}^N \CP_{\pd \Delta_i}\Big)\Big/\!\!\!\Big/(\C^*)^{N-d}\,. \label{3d:sympred} \ee
The $N-d$ moment maps $\mu_I$ in the symplectic reduction are simply the products of tetrahedron edge-coordinates $z_i,z_i',z_i''$ around every internal edge $E_I$ created in the gluing. Fixing $\mu_I=1$ ensures that a classical flat connection is smooth at that edge. In addition, every $\C^*$ coordinate in $\CP_2(\pd M)$ is expressed as a Laurent monomial in tetrahedron edge-coordinates (well defined up to multiplication by the $\mu_I$). For example, every $x_E$ on the big boundary of $M$ is a product of the tetrahedron edge-coordinates incident to the edge $E$.

The Lagrangian $\CL_2(M)\subset \CP_2(\pd M)$ can also be obtained%
\footnote{Strictly speaking, this is true only for a sufficiently generic or refined triangulation of $M$. In particular, one must make sure that the $(\C^*)^{N-d}$ action in the quotient is transverse to the product Lagrangian $\prod_i \CL_{\Delta_i}$.} %
 by ``pulling'' a canonical product Lagrangian $\prod_i \CL_{\Delta_i}\subset \prod_i \CP_{\pd \Delta_i}$ through the symplectic reduction \eqref{3d:sympred}. This means projecting $\prod_i \CL_{\Delta_i}$ along the $(\C^*)^{N-d}$ flows of the moment maps $\mu_I$, and intersecting with the locus $\mu_I=1$. This gives a very hands-on algebraic construction of a moduli space that otherwise may appear extremely complicated.

It is known how to generalize the symplectic-gluing construction of $\CL_K(M)\subset \CP_K(\pd M)$ to arbitrary $K$. Moreover, it is straightforward to quantize the entire construction \cite{Dimofte-QRS}. Combinatorially, quantization requires taking logarithms of all cross-ratio coordinates, and consistently keeping track of their imaginary parts. This corresponds physically to keeping track of the $U(1)_R$ symmetry of $T_K[M]$ on curved backgrounds.


\subsubsection{Limitations}
\label{3d:sec:sub}

We have noted in passing that when we construct Lagrangian $\CL_K[M]$ from tetrahedra by symplectic gluing, we may only recover an algebraically open patch of the full moduli space of framed flat connections on $M$.
The basic limitation is that all cross-ratio coordinates $z,z',z''$ for tetrahedra in a triangulation of $M$ must be non-degenerate: not equal to $0$, $1$, or $\infty$. Equivalently, the four framing flags at the vertices of any tetrahedron must be distinct after parallel transport to the center.
This restriction can sometimes cause the glued Lagrangian $\CL_K[M]$ to miss entire families of flat connections.
Then, if we use an analogous gluing construction to build a 3d $\CN=2$ theory, as in the next section, we may only recover a subsector
of the full $T_K[M]$, whose vacua on $S^1$ correspond only to \emph{some} of the flat connections on $M$. This was recently emphasized in \cite{CDGS}.

To illustrate what we mean in terms of flat connections, suppose that $M$ is a knot complement, \ie\ $S^3$ with a knotted defect inside, which has been regularized to a small torus boundary. A flat $SL(2,\C)$ connection on $M$ induces (via its holonomies) a representation $\rho:\pi_1(M)\to SL(2,\C)$, and can be classified by the ``reducibility'' of this representation, \ie\ the subgroup of $SL(2,\C)$ that commutes with the image $\rho(\pi_1(M))$. For example, only the identity element commutes with a fully irreducible representation, while a full $GL(1)\subset SL(2,\C)$ commutes with an ``abelian'' representation (whose holonomies can all be simultaneously diagonalized). Typically both types of representations exist: there is always an abelian representation, while for hyperbolic knot complements the holonomy of the hyperbolic metric is always irreducible.
If we now choose a triangulation for $M$ and choose a framing line on $\pd M=T^2$, we find that all vertices of all tetrahedra share the same framing line (since all vertices land on the same $T^2$), and the only way to get non-degenerate cross-ratios is to have non-trivial parallel transport inside the tetrahedra. However, the parallel transport of an abelian flat connection acts trivially on the framing lines --- and tetrahedron cross-ratios for an abelian flat connection are always degenerate. Therefore, only non-abelian representations are captured by symplectic gluing of tetrahedra.

This is not a serious problem when $K=2$ and all components of $\pd M$ have genus $>1$, such as for manifolds encoding duality domain walls in theories $T_2[C]$ of class $\CS$, when $C$ has negative Euler character. In this case, generic choices of boundary conditions (eigenvalues of boundary holonomies) completely forbid reducible flat connections on $M$. For example, the manifold in Figure \ref{3d:fig:wallgeoms}a, encoding the S-duality wall for $\CN=2^*$ theory, has a total boundary of genus $2$.  Then triangulation methods readily reconstruct $T_2[M] \simeq T[SU(2)]$, without missing any branches of vacua \cite{DGV-hybrid}.

In higher rank $(K\geq 3)$ the issue is more severe. Non-degeneracy of cross-ratios requires all the defects in a manifold $M$ to be of ``maximal'' type, carrying maximal $SU(K)$ flavor symmetry (so that all eigenvalues of boundary holonomies can be distinct). Subsequently, only fully irreducible flat connections are captured by the standard symplectic gluing of~\cite{DGG-Kdec}.

The precise physical significance of the subsector of $T_K[M]$ coming from gluing tetrahedra is still being elucidated. Thinking of $T_K[M]$ as the theory of $K$ M5 branes wrapping $M\times \R^3\subset T^*M\times\R^3\times\R^2$, as in Section \ref{3d:sec:6d}, a plausible conjecture is that the subsector obtained by gluing tetrahedra only captures the physics of configurations where the $K$ M5's reconnect into a \emph{single} M5 wrapping a spectral cover of $M$. Thus the subsector is missing configurations where the $K$ M5's reconnect into multiple components (or remain fully disconnected), and are thus able to separate in the $\R^2$ direction. Such configurations would correspond to the missing branches of vacua. This conjecture is in line with findings of \cite{CDGS}, where it was argued in examples that the full $T_K[M]$ contains an additional $U(1)_t$ flavor symmetry, involving rotations of $\R^2$.

\subsection{The tetrahedron theories}
\label{3d:sec:TD}

Just as framed 3-manifolds are glued together from tetrahedra, the 3-manifold theories $T_K[M]$ (or more precisely $T_K[M,\Pi]$ or $T_K[M,\mb p]$) are glued together from tetrahedron theories. For simplicity, we will review how this works in the case $K=2$.

The first step is to identify the theory of a single truncated tetrahedron. As we first tried to motivate physically in Section \ref{3d:sec:6d}, however, there should be no \emph{unique} tetrahedron theory. Rather, there is an infinite family of 3d theories $T_2[\Delta,\Pi]$ labelled by choices of polarization $\Pi$ on the boundary of the tetrahedron --- \emph{a.k.a.} ways of decoupling an abelian 4d bulk gauge theory from a 3d boundary condition. Now we can understand the polarization in a purely geometric setting: $\Pi$ is a choice of ``electric'' $\C^*$ position coordinate and canonically conjugate ``magnetic'' $\C^*$ momentum coordinate for $\CP_{\pd \Delta}$.

Choosing
\be \Pi = \Pi_z := \begin{pmatrix} \text{position}=z \\\text{momentum}=z'' \end{pmatrix}\,, \label{3d:Pz} \ee
with canonical Poisson bracket $\{\log z'',\log z\}=1$, the tetrahedron theory was conjectured in \cite{DGG} to be
\be T_\Delta := T_2[\Delta,\Pi_z] = \left\{ \begin{array}{l} \text{free chiral $\phi_z$ with $U(1)_z$ flavor symmetry}\,; \\ \text{background CS level $-1/2$ for $U(1)_z$}\,.
\end{array}\right.  \label{3d:TD} \ee
This agrees beautifully%
\footnote{Note that the half-integer background Chern-Simons term is corrected by the standard parity anomaly of a 3d $\CN=2$ theory (\cf\ \cite{AHISS}) to be an integer in the IR, given any nonzero real mass for $\phi_z$.} %
with the theory intuited from an analysis of the tetrahedron's spectral cover in Section \ref{3d:sec:SWwall}.

The symplectic group $Sp(2,\Z)$ acts both on a formal polarization vector such as \eqref{3d:Pz} and on a 3d SCFT with a $U(1)$ flavor symmetry, as described in \cite{Witten-sl2}. The provides a concrete way to change the polarization of a theory; for example, we expect
\be T_2[\Delta,g\circ \Pi_z] = g\circ T_2[\Delta,\Pi_z]\,,\qquad g\in Sp(2,\Z)\,. \ee
Concretely, the generator $T={\large\left(\begin{smallmatrix} 1&0\\1&1 \end{smallmatrix}\right)}$ acts on a theory by adding $+1$ to the background Chern-Simons level for the flavor symmetry. The generator $S={\large\left(\begin{smallmatrix} 0&-1\\1&0 \end{smallmatrix}\right)}$ gauges the flavor $U(1)$, after which there appears a new ``topological'' flavor symmetry $U(1)_J$. These actions can be understood as the effect of electric-magnetic duality on the 3d boundary of a 4d abelian gauge theory.

Although we can choose any polarization we want for the tetrahedron theory, three of them are special: the polarizations in which one of the edge-coordinates themselves (\ie\ $z$, $z'$, or $z''$ rather than an arbitrary Laurent monomial like $z^3z'{}^{-1}$) is a position. We can call these $\Pi_z$, $\Pi_{z'}$, and $\Pi_{z''}$. In fact, since the cyclic rotation symmetry of the tetrahedron permutes $z\to z'\to z''\to z$, we might even expect that the resulting theories are all equivalent:
\be \label{3d:TDequiv}
 T_2[\Delta,\Pi_z] \simeq T_2[\Delta,\Pi_{z'}] \simeq T_2[\Delta,\Pi_{z''}]\,. \ee
This is indeed true. For example, to pass from $\Pi_z$ to $\Pi_{z'}$, we act with $ST\in Sp(2,\Z)$,
\be \Pi_{z'} = \begin{pmatrix} z' \\ z \end{pmatrix} = \begin{pmatrix} -\frac{1}{zz''} \\ z \end{pmatrix} = \begin{pmatrix} -1 & -1 \\ 1 & 0 \end{pmatrix}\cdot  \begin{pmatrix} z \\ z'' \end{pmatrix} = ST\circ \Pi_z\,, \ee
where the linear transformation acts \emph{multiplicatively} (\ie\ $\large \left(\begin{smallmatrix} a& b\\c & d\end{smallmatrix}\right)\cdot
\left(\begin{smallmatrix} z\\ w\end{smallmatrix}\right) = 
\left(\begin{smallmatrix} z^aw^b\\ z^c w^d\end{smallmatrix}\right)$\,), and we are ignoring signs%
\footnote{The signs, and indeed the full lift to logarithms of the edge-coordinates, becomes relevant when keeping track of a choice of $U(1)_R$ symmetry for a theory. Then symplectic $Sp(2N,\Z)$ actions are promoted to affine-symplectic $ISp(2N,\Z)$ actions.} %
such as $(-1)\frac 1{zz''}$. Correspondingly, we find
\be \label{3d:TDp} T_2[\Delta,\Pi_{z'}] = ST\circ T_2[\Delta,\Pi_z]= \left\{\begin{array}{l}
 \text{$U(1)$ gauge theory with chiral $\phi_{z'}$ of charge +1}\,;\\
 \text{CS level +1/2 for the dynamical $U(1)$}\,; \\
 \text{topological $U(1)_{z'}$ flavor symemtry}\,. \end{array}\right.
\ee
In the infrared, this theory flows to the \emph{same} SCFT $T_\Delta$ as in \eqref{3d:TD}. The monopole operator of \eqref{3d:TDp} (which creates free vortices) matches the free chiral of \eqref{3d:TD} \cite{AHISS, DGG}. This match is strong evidence that the tetrahedron theory has been properly identified.

Yet another piece of evidence that \eqref{3d:TD} is correct comes from compactifying the theory on a circle $S^1$ and calculating its supersymmetric parameter space \eqref{3d:LSUSY}. A straightforward summation of Kaluza-Klein modes (\cf\ \cite{NS-I}) leads to the twisted superpotential $\wt W(z) = \Li_2(z^{-1})$, where $\log z$ is the complexified mass associated to the $U(1)_z$ flavor symmetry. Then the definition of the effective FI parameter
\be \exp \frac{\pd \wt W(z)}{\pd z} = z''\quad\Rightarrow\quad z''+z^{-1}-1=0 \ee
reproduces the tetrahedron Lagrangian $\CL_\Delta$ from \eqref{3d:LD}, as desired.

\subsection{Gluing together theories}
\label{3d:sec:glueT}

Now suppose that a framed 3-manifold $M$ is glued together from $N$ tetrahedra. In order to define an isolated 3d theory $T_2[M,\Pi]$, we need to choose a polarization $\Pi$ for the big boundary of $M$,%
\footnote{In Section \ref{3d:sec:6d}, we also talked about isolating 3d theories $T_K[M,\mb p]$ based on a pants decomposition $\mb p$ of the topological boundary of $M$. This was meant to correspond to decoupling a nonabelian 4d gauge theory in some duality frame. Such a choice is already \emph{built in} to the definition of a \emph{framed} manifold $M$: a pants decomposition for a boundary component $\CC$ corresponds to a splitting of that boundary into a network of small annuli connected by big 3-punctured spheres when selecting a framing.} %
or rather for the part of $\CP_2(\pd M)$ corresponding to the big boundary. For any small tori in $\pd M$, we also choose A- and B-cycles. For small annuli, though, the choice of non-contractible ``A-cycles'' (and so the polarization) is canonical.

We build $T_2[M,\Pi]$ by first taking a ``tensor product'' of tetrahedron theories
\be  T_\times = T_{\Delta_1}\times \cdots \times T_{\Delta_N}\,,\ee
which is basically a collection of $N$ free chirals $\phi_{z_i}$ with flavor symmetry $\prod_iU(1)_{z_i}\simeq U(1)^N$. This product theory corresponds to a product polarization $\Pi_\times =(\text{positions $z_i$; momenta $z_i''$})$ on the product phase space $\prod_i \CP_{\pd \Delta_i}$.

Now the symplectic group $Sp(2N,\Z)$ acts to change the polarization of $T_\times$. This is a natural extension of the $Sp(2,\Z)$ action on theories with a single $U(1)$ symmetry: the action of an element $g\in Sp(2N,\Z)$ just modifies various CS levels, gauges some of the $U(1)$'s in $U(1)^N$, and/or permutes the $U(1)$ factors in $U(1)^N$.

We then choose a \emph{new} polarization $\wt \Pi_\times=g\circ \Pi_\times$ for $T_\times$, determined by the following algebraic properties:
\begin{enumerate}
\item all the position and momentum coordinates of $\Pi$ (as monomial functions on $\prod_i \CP_{\pd \Delta_i}$) are positions and momenta, respectively, in $\wt \Pi_\times$; and
\item all the moment maps $\mu_I$ (products of tetrahedron edge-coordinates around internal edges in $M$) are positions in $\wt \Pi_\times$.
\end{enumerate}
The first requirement simply makes $\wt \Pi_\times$ compatible with our desired final polarization $\Pi$. The second requirement, however, is absolutely crucial for the gluing: it guarantees%
\footnote{Just like in the gluing of classical Lagrangian submanifolds, some extra regularity conditions need to be imposed on a 3d triangulation to truly guarantee the existence of the gluing operators $\CO_I$. See Section 4.1 of \cite{DGG} or the Appendix A of \cite{DGG-Kdec}.} %
that the transformed product theory $g\circ T_\times$ will contain \emph{chiral operators} $\CO_I$ associated to each internal edge $E_I$ of $M$. Each of these operators $\CO_I$ will transform under a flavor symmetry associated to the internal-edge coordinate $\mu_I$.

The final step in the gluing is to add the $N-d$ internal-edge operators $\CO_I$ to the superpotential of $g\circ T_\times$. This breaks $N-d$ $U(1)$ flavor symmetries, and implements the symplectic reduction \eqref{3d:sympred} on the gauge-theory level. The result is a UV abelian Chern-Simons-matter theory with manifest $U(1)^{d}$ flavor symmetry, which flows in the IR to $T_2[M,\Pi]$.

\section{Examples}

We finish with a brief look at two simple framed 3-manifolds $M$ and their effective theories at $K=2$. We'll mainly follow the bottom-up approach of symplectic gluing from tetrahedra; though both examples are amenable to top-down analyses as well.

\begin{figure}
\centering
\includegraphics[width=4.3in]{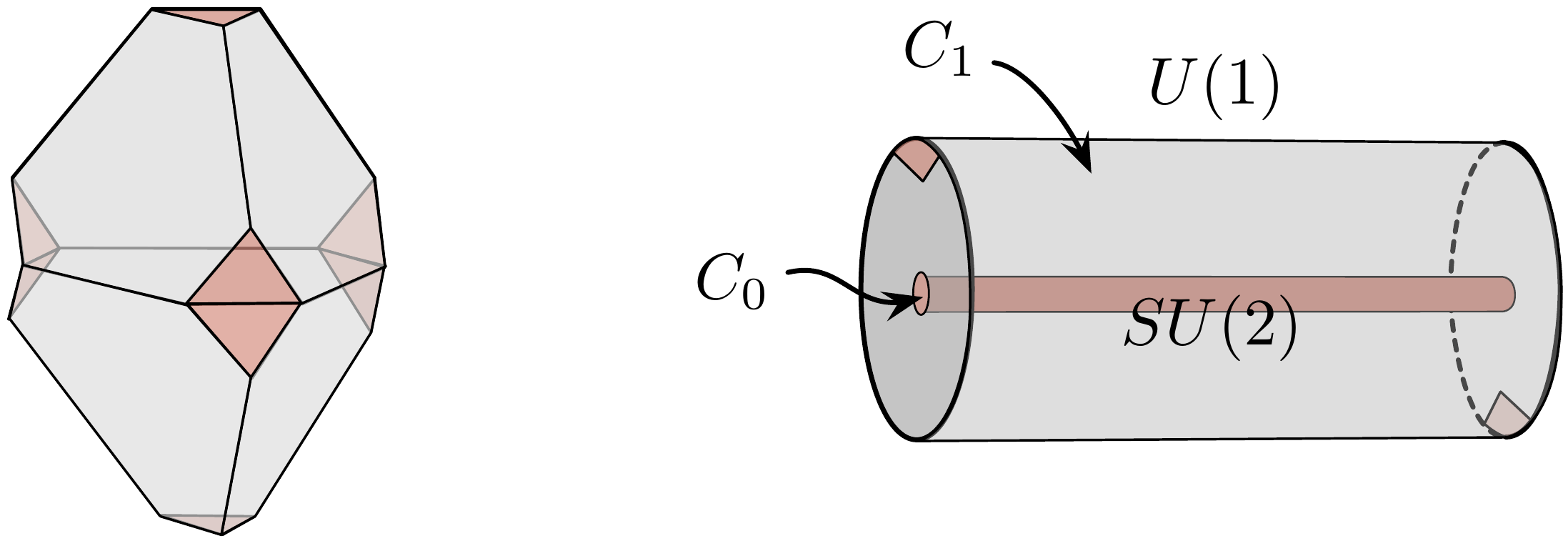}
\caption{\small The bipyramid (left) and the thickened annulus representing the RG manifold (right).}
\label{3d:fig:23RG}
\end{figure}

The first example, introduced in \cite{DGG}, is a triangular bipyramid (Figure \ref{3d:fig:23RG}, left). Like a truncated tetrahedron, it only has disc-like small boundaries (at the five truncated vertices), and a big boundary consisting of a five-holed sphere. The bipyramid can be assembled from gluing either two or three tetrahedra together. The IR equivalence of the glued theories that result (containing either two or three chirals) provides the local proof of triangulation independence for general glued theories $T_2[M,\Pi]$ (in fact also for $K>2$).

The second example is a 3-manifold with topology $M=C\times I$, where $C$ is a cylinder and $I=\{0\leq t\leq 1\}$ an interval. We picture $M$ as a solid cylinder with a core drilled out (Figure \ref{3d:fig:23RG}, right). To specify $M$ as a \emph{framed} 3-manifold, we take the boundary $C_0$ at $t=0$ (the core in the solid-cylinder picture)
 to be a small annulus. The remainder of $\pd M$ is split into a big annulus $C_1$, glued to two big punctured discs (the ends of the solid cylinder, $\pd C\times I$), with two additional small discs sandwiched inbetween (drawn as tiny triangular regions in Figure \ref{3d:fig:23RG}). Thus, topologically, total full big boundary of $M$ is a 4-holed sphere.
This manifold turns out to be the basic building block of RG domain walls, as well as more general UV S-duality walls, as discussed in Section \ref{3d:sec:duality} (and in great detail in \cite{DGV-hybrid}). Geometrically, $M$ represents the local shrinking of an annular region on any surface to a long, thin tube, and ultimately to a defect. We will see that the theory $T_2[M,\Pi]$ has $SU(2)\times U(1)$ flavor symmetry, allowing a coupling to a nonabelian 4d gauge group on one side, and an abelian gauge group on the other.

\subsection{2--3 move and mirror symmetry}

\begin{figure}[htb]
\centering
\includegraphics[width=5.7in]{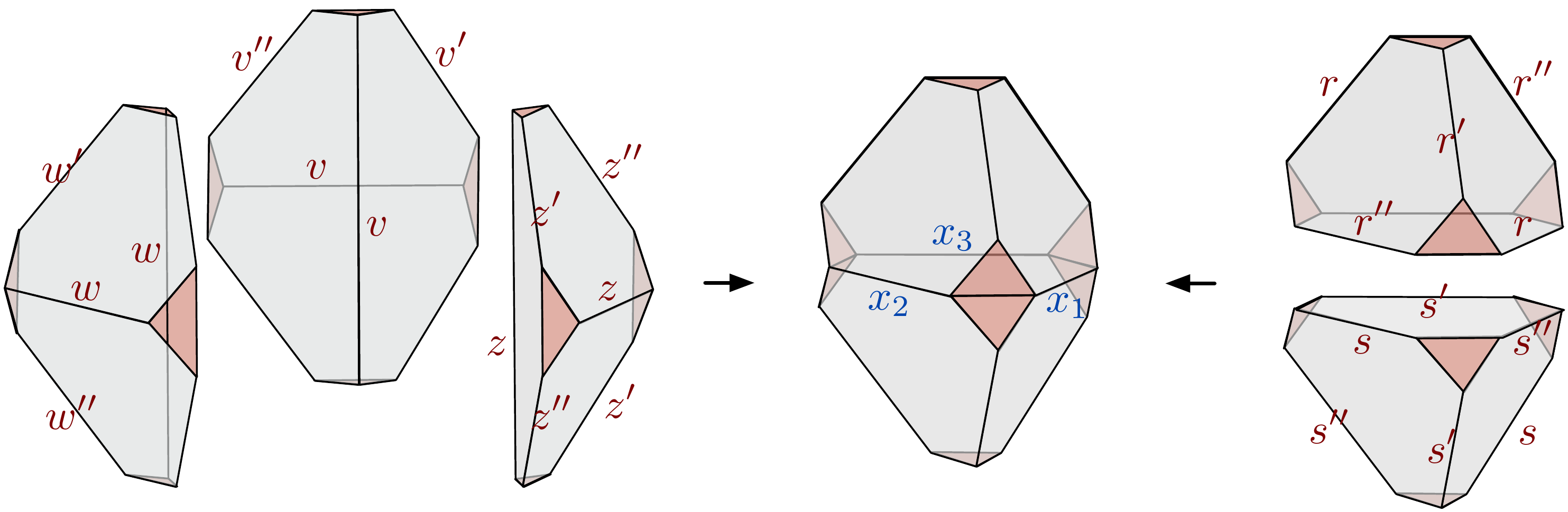}
\caption{\small Gluing together the bipyramid from two or three tetrahedra.}
\label{3d:fig:23}
\end{figure}

Let $M$ be the triangular bipyramid. Let's first observe that $M$ has a boundary phase space $\CP_2(\pd M)\simeq (\C^*)^4$. It is easy to see this: one can construct cross-ratio coordinates $x_E$ for each of the nine edges on the boundary, while each of the five vertices imposes a relation that the product of edge-coordinates around that vertex equals $\pm 1$ (for unipotent holonomy). Thus $\dim_\C \CP_2(\pd M)=9-5=4$. We will choose a polarization $\Pi_{\rm eq}$ for $\CP_2(\pd M)$ such that two of the three equatorial edges of the bipyramid $(x_1,x_2)$ carry electric/position coordinates, as in the center of Figure \ref{3d:fig:23}. Since the product of all equatorial edges is one, this implies that the third edge $x_3=x_1^{-1}x_2^{-1}$ is electric or ``mutually local'' as well. Note that specifying the position (but not momentum) coordinates in a polarization is sufficient to define a theory $T_2[M,\Pi_{\rm eq}]$ up to background Chern-Simons levels.

Now, suppose that we glue together a bipyramid $M$ from three tetrahedra, as on the LHS of Figure \ref{3d:fig:23}. We must polarize the tetrahedra, and we choose standard polarizations \eqref{3d:Pz}, in such a way that the unprimed position coordinates $z,w,v$ all lie along the internal edge of $M$. Now the three equatorial edges on the boundary of the bipyramid also get coordinates $z,w,v$ (from opposite edges of the three tetrahedra). So no change of polarization is needed to make the product polarization $\Pi_z\times\Pi_w\times \Pi_v$ on the tetrahedra compatible with our final desired $\Pi_{\rm eq}$. The bipyramid theory $T_2[M,\Pi_{\rm eq}]$ is then easy to write down: it is just the product $T_{\Delta z}\times T_{\Delta w}\times T_{\Delta v}$ containing three chirals $\phi_z,\phi_w,\phi_v$, in which the $U(1)^3$ flavor symmetry is broken to $U(1)^2$ by a cubic superpotential
\be \CO_I = \phi_z\phi_w\phi_v \ee
corresponding to the internal edge. This theory is usually called the ``XYZ model.'' Note how the individual operators $\phi_z,\phi_w,\phi_v$ are each associated to one of the electric edges on~$\pd M$.

Let us also explain the symplectic reduction geometrically.
We can explicitly write the boundary phase space as
\begin{align} \CP_2(\pd M) &= \big(\CP_{\pd \Delta z} \times \CP_{\pd \Delta z} \times \CP_{\pd \Delta z}\big)\big/\!\!\big/ \C^* \notag\\ 
& \simeq \{z,z'',w,w'',v,v''\in \C^*\}\big/\raisebox{-.1cm}{\small $(z'',w'',v'')\sim (tz'',tw'',tv'')$}\,\big|\raisebox{-.1cm}{\small $zwv=1$}\,,
\end{align}
where we have quotiented with respect to the flows of the moment map $\mu_I = zwv$, and intersected with the locus $\mu_I=1$. Notice that all products of tetrahedron coordinates on external edges (such as $z,w,v,$ or $z'w'',w'v''$, etc.), \emph{commute} with $\mu_I$, and so form good coordinates $x_E$ on the quotient. (For a computation of the Lagrangian submanifold $\CL_2(M)$ and its quantization, see \cite{Dimofte-QRS} or \cite{DGG}.)

Alternatively, if we form the bipyramid from two tetrahedra, there are no internal edges created, but a nontrivial change of polarization is required. Let us assign triples of coordinates to the tetrahedra as on the RHS of Figure \ref{3d:fig:23}, and choose standard polarizations $\Pi_r$, $\Pi_s$ for them. The equatorial coordinates for the bipyramid are related to tetrahedron coordinates as
\be x_1 = rs''\,,\quad x_2=r''s\,,\qquad \big( x_3 = r's'= (rr''ss'')^{-1}=(x_1x_2)^{-1}\big)\,,\ee
and so involve both tetrahedron positions $(r,s)$ and momenta $(r'',s'')$. The $Sp(4,\Z)$ change of polarization that relates $\Pi_r\times \Pi_s$ to $\Pi_{\rm eq}$ acts on the theory $T_{\Delta r}\times T_{\Delta s}$ by gauging%
\footnote{The precise $Sp(4,\Z)$ action first removes the background Chern-Simons coupling for the anti-diagonal subgroup of $U(1)_r\times U(1)_s$, and then gauges it. It is a nice exercise to demonstrate this.} %
the anti-diagonal subgroup of the flavor symmetry group $U(1)_r\times U(1)_s$. The resulting theory is just 3d $\CN=2$ SQED, which is mirror symmetric to the XYZ model \cite{AHISS}.
It has an axial $U(1)_{ax}$ and a topological $U(1)_J$ flavor symmetry, matching the $U(1)^2$ flavor symmetry of the XYZ model. Moreover, it has monopole and anti-monopole operators $\eta_\pm$ in addition to the gauge-invariant meson $\varphi=\phi_r\phi_s$, which together match the three chiral operators $\phi_z,\phi_w,\phi_v$ of the XYZ model, and label the equatorial edges of the bipyramid.

\subsection{The basic RG wall}

\begin{figure}[htb]
\centering
\includegraphics[width=5in]{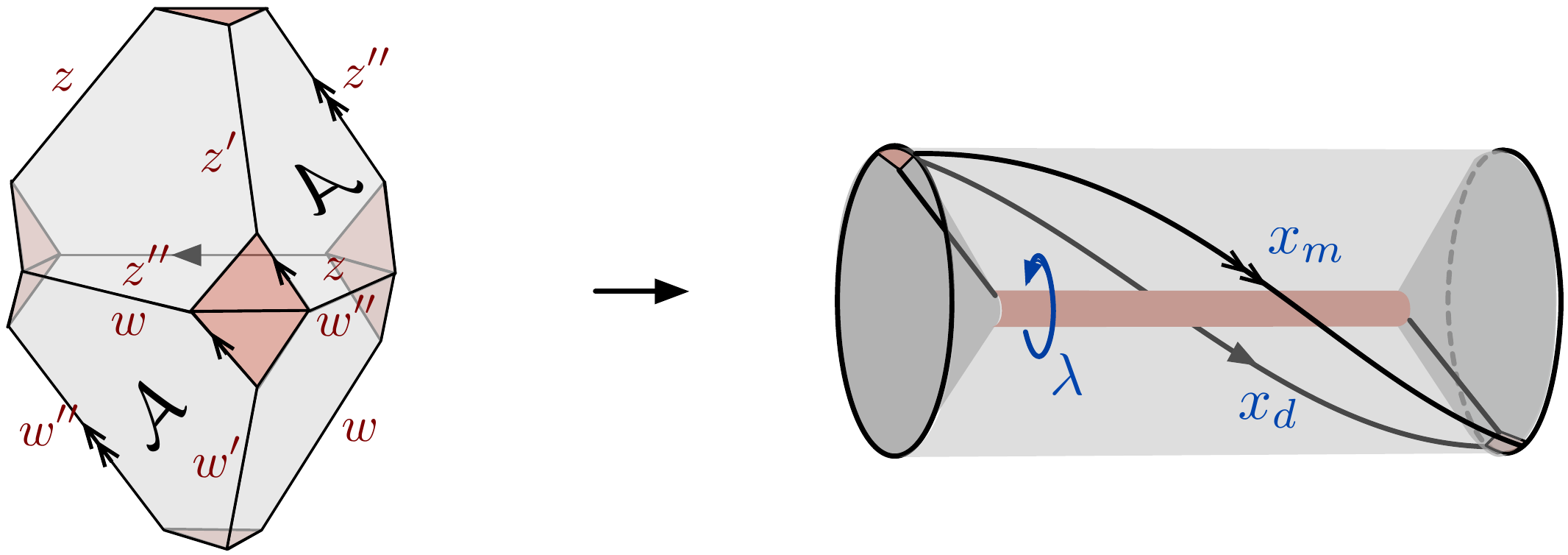}
\caption{\small Forming the RG-wall manifold $M$ by identifying two faces of the bipyramid, as indicated by labels `$\CA$' on the left. On the right, we show the triangulation on the big boundary of~$M$.}
\label{3d:fig:RG}
\end{figure}

Now let $M$ be the RG-wall manifold. Just like the bipyramid, it also has a 4-complex dimensional phase space. Independent coordinates on $\CP_2(M)$ are now given by cross-ratios $(x_m,x_d)$ on two edges of the big annulus $C_1$ (compare Figures \ref{3d:fig:23RG} and \ref{3d:fig:RG}) together with an eigenvalue $\lambda$ of the $PSL(2)$ holonomy%
\footnote{Two technical clarifications here: first, the choice of eigenvalue $\lambda$ vs. $\lambda^{-1}$ depends on the choice of framing for the flat connection at the small annulus; second, to get a well defined sign for $\lambda$ one actually needs to lift to $SL(2)$ rather than $PSL(2)$ holonomies aroudn the small annulus.} %
around the girth of the small annulus $C_0$ and its canonical conjugate, a twist coordinate $\tau$:
\be \CP_2(M)\simeq \{x_m,x_d,\lambda,\tau\}\simeq (\C^*)^4\,, \ee \vspace{-.8cm}
\be \{\log x_d,\log x_m\}= 2\,,\quad \{\log \tau,\log \lambda\}=1\,,\qquad\text{other brackets vanishing}. \notag \ee
We will choose a polarization $\Pi_e$ with position coordinates $\lambda$ and $x_e = (x_mx_d)^{-1/2}$.

We can build $M$ from two truncated tetrahedra, as shown in Figure \ref{3d:fig:RG}. There are no internal edges, so no superpotentials will be needed. We give the tetrahedra edge-coordinates $z,z',z''$ and $w,w',w''$ and standard polarizations $\Pi_z,\Pi_w$. Then we find
\be \lambda = \sqrt{\frac zw}\,,\qquad x_e = \sqrt{zw}\,\ee
(as well as $x_m=z''w'',\,x_d=z'w',\,\tau=\lambda z''/w''$). Since $\lambda$ and $x_e$ are just made from tetrahedron positions $z,w$, the change of polarization $\Pi_z\times \Pi_w \to \Pi_e$ involves no gauging, just a redefinition of flavor symmetries. We find that $T_2[M,\Pi_e]$ is a theory of two free chirals $\phi_z,\phi_w$ transforming with charges $(+1,-1)$ and $(+1,+1)$, respectively, under $U(1)_\lambda$ and $U(1)_e$ flavor symmetries associated to $\lambda$ and $x_e$. Of course the vector $U(1)_\lambda$ symmetry is actually enhanced to $SU(2)_\lambda$. As promised, the extremely simple theory $T_2[M,\Pi_e]$ can couple both to $SU(2)$ and $U(1)$ 4d gauge groups.

Alternatively, had we chosen a polarization $\Pi_m$ with $\lambda$ and $x_m$ as positions, we would instead have described $T_2[M,\Pi_m]$ as a theory of two chirals $\phi_z,\phi_w$ whose axial $U(1)_e$ symmetry is gauged at Chern-Simons level $-1$, and replaced by a topological $U(1)_m$. This is roughly the UV GLSM description of a 3d $\cp^1$ sigma model. Now the theory has a monopole operator $\CO_m$ associated to the external ``electric'' edge with coordinate $x_m$. Similarly, we could have chosen a polarization $\Pi_d$ to obtain a theory $T_2[M,\Pi_d]$ whose axial $U(1)_e$ is gauged at Chern-Simons level $+1$.

The claim of \cite{DGV-hybrid}, a full review of which is beyond our scope, is that the theories $T_2[M,*]$ are effective theories for an RG domain wall in pure $SU(2)$ Seiberg-Witten theory. In the respective polarizations $\Pi_e,\Pi_m,\Pi_d$, the 3d theories couple to the abelian 4d theory on its Coulomb branch --- in 4d duality frames so that the electric, magnetic, or dyonic gauge fields are fundamental. In all these polarizations, the 3d theory couples on the other side of the wall to the nonabelian UV gauge group $SU(2)_\lambda$.

One way to create an RG wall in pure $SU(2)$ theory is by engineering a Janus configuration (\cf\ \eqref{3d:Janus}) where the UV cutoff $\Lambda$ varies (relative to a fixed observation scale) as a function of the space coordinate $x^3$. To the left of the wall, $\Lambda$ can be arbitrarily close to zero, effectively putting the 4d theory in the UV; while to the right of the wall $\Lambda$ can be sent close to infinity. We observe the theory at an intermediate energy scale throughout. This traps 3d degrees of freedom on the wall. We can even make an educated guess at what they should be.

Passing through the wall from left to right, the imaginary part of $a(x^3)$ is forced to infinity (relative to our observation scale), breaking $SU(2)\to U(1)$ and Higgsing the 4d theory. However, close to the (left of the) wall, the $SU(2)$ gauge fields are effectively non-dynamical, since the gauge coupling is infinitesimally small. Thus Goldstone bosons cannot be eaten up by $W$-bosons, and parametrize a $\cp^1\simeq SU(2)/U(1)$ -worth of degrees of freedom at the wall. This beautifully matches the bottom-up constructions of $T_2[M,*]$.

The RG walls (and nonabelian S-duality walls) of more complicated 4d theories always involve components that look like the theories $T_2[M,*]$. Indeed, whenever one has a framed 3-manifold $\wh M$ with a network of small annuli connecting big boundaries, the neighborhood of every small annulus can be made to look exactly like our RG-manifold $M$. This proves, among other things, that in a bottom-up construction of $T_2[\wh M]$, all the $U(1)$ symmetries associated to small annuli will be enhanced to $SU(2)$'s --- as must be the case if the small annuli are to represent defects in a 6d compactification.

Finally, let us see what information is contained in the Lagrangian submanifold $\CL_2(M)$ of the RG-wall manifold. By rewriting the tetrahedron Lagrangians $z''+z^{-1}-1=w''+w^{-1}-1=0$ in terms of $x_m,x_d,\lambda,\tau$ and $x_e=1/\sqrt{x_mx_d}$, we find
\begin{subequations} \label{3d:RGL}
\begin{align} \big(\text{Wilson}_{\frac12}\big)&\qquad \lambda+\lambda^{-1} 
= x_e+x_e^{-1}-x_ex_m \hspace{.5in} \\
\big(\text{'t Hooft}_{\frac12}\big)&\qquad \frac{(\tau\lambda)^{\frac12}-(\tau\lambda)^{-\frac12}}{\lambda-\lambda^{-1}} = \frac{1}{\sqrt{x_m}} \\
\big(\text{'t Hooft-Wilson}_{\frac12}\big)& \qquad \frac{(\tau/\lambda)^{\frac12}-(\tau/\lambda)^{-\frac12}}{\lambda-\lambda^{-1}} = \sqrt{x_d}\,. \end{align}
\end{subequations}
The first equation relates the spin-1/2 UV Wilson line of pure $SU(2)$ Seiberg-Witten theory to IR line operators of abelian electric and magnetic charge \cite{GMNIII, DG-Sdual}. The second and third equations (which are not independent) relate the spin-1/2 UV 't Hooft lines and mixed 't Hooft-Wilson lines to the IR magnetic and dyonic line operators. The honest $SU(2)$ theory should only contain magnetic UV operators of spin-one, corresponding (roughly) to squaring equations \eqref{3d:RGL}b-c, which then gets rid of the square roots. The quantization of relations \eqref{3d:RGL} turns out to match operator equations known from quantum Teichm\"uller theory on the annulus \cite{Kash-kernel, Teschner-TeichMod}, giving a beautiful geometric interpretation of the latter.

\subsection*{Acknowledgements}
It is a pleasure to thank Christopher Beem, Clay C\'ordova, Davide Gaiotto, and Sergei Gukov for discussions and advice during the writing of this review, and especially Andrew Neitzke and Jeorg Teschner for careful readings and comments.

\bibliographystyle{JHEP_TD}
\bibliography{toolbox}

\end{document}